\documentclass[preprint,12pt]{elsarticle}

\usepackage{amssymb}
\usepackage{amsthm}
\usepackage{xcolor}
\usepackage{graphicx}
\usepackage{lineno,hyperref}
\modulolinenumbers[5]

\usepackage{inputenc}
\usepackage{amsmath}
\usepackage{graphics}
\usepackage{epsfig}
\usepackage{graphicx}
\usepackage{latexsym}
\usepackage{graphics}
\usepackage{epsfig}
\usepackage{amsmath,amssymb,amsfonts,theorem,color,bm}
\usepackage{multirow}
\usepackage{stfloats} 
\usepackage{float}
\usepackage{subfig}
\usepackage{soul}
\usepackage{caption}
\usepackage{hyperref}
\hypersetup{colorlinks,allcolors=black}

\newcommand\bc{\boldsymbol c}

\newcommand\by{\boldsymbol y}

\newcommand\bQ{\boldsymbol Q}

\newcommand\bv{\boldsymbol v}
\newcommand\bV{\boldsymbol V}

\newcommand\bW{\boldsymbol W}

\newcommand\bC{\boldsymbol C}

\newcommand\bT{\boldsymbol T}

\newcommand\bPhi{\boldsymbol{\Phi}}
\newcommand\bSigma{\boldsymbol{\Sigma}}

\newcommand\bR{\boldsymbol{R}}

\numberwithin{equation}{section}

\newcommand{\beqn}{\begin{equation}}
\newcommand{\eeqn}{\end{equation}}
\newcommand{\beqnarr}{\begin{eqnarray}}
\newcommand{\eeqnarr}{\end{eqnarray}}
\newcommand{\baling}{\begin{alignat}{1}}
\newcommand{\ealing}{\end{alignat}}

\usepackage{xcolor,colortbl}
\definecolor{Gray}{gray}{0.75}
\newcolumntype{a}{>{\columncolor{Gray}}c}

\journal{XXX}

\begin{document}

\begin{frontmatter}



\title{Low-cost singular value decomposition with optimal sensor placement}


\author[UPM]{Ashton Hetherington}


\author[UPM]{Soledad Le Clainche\footnote{Correspondence to: soledad.leclainche@upm.es}}

\affiliation[UPM]{organization={ETSI Aeronáutica y del Espacio, Universidad Politécnica de Madrid},
            addressline={Plaza Cardenal Cisneros, 3}, 
            city={Madrid},
            postcode={28040}, 
            country={Spain}}

\begin{abstract}
This paper presents a new method capable of reconstructing datasets with great precision and very low computational cost using a novel variant of the singular value decomposition (SVD) algorithm that has been named low-cost SVD (lcSVD). This algorithm allows to reconstruct a dataset from a minimum amount of points, that can be selected randomly, equidistantly or can be calculated using the optimal sensor placement functionality that is also presented in this paper, which finds minimizing the reconstruction error to validate the calculated sensor positions. This method also allows to find the optimal number of sensors, aiding users in optimizing experimental data recollection. The method is tested in a series of datasets, which vary between experimental and numerical simulations, two- and three-dimensional data and laminar and turbulent flow, which have been used to demonstrate the capacity of this method based on its high reconstruction accuracy, robustness, and computational resource optimization. Maximum speed-up factors of 630 and memory reduction of 37\% are found when compared to the application of standard SVD to the dataset. This method will be incorporated into ModelFLOWs-app's\footnote{The website of the software is available at \href{https://modelflows.github.io/modelflowsapp/}{https://modelflows.github.io/modelflowsapp/}} next version release.
\end{abstract}




\end{frontmatter}



\section{Introduction\label{sec:introduction}}
One of the main obstacles encountered when working fluid mechanics databases is their large size. In some cases, users have the urge to search for alternative algorithms which allow them to work with these databases at a low computational cost.

Some of these solutions include high performance computing techniques, such as parallelization, which allow users to spread out the work load over the available processors or GPU units. Just to name a few examples, parallelization can be used to speed-up direct numerical simulation solvers in Python \cite{mortensen2016high}, or to employ non-linear optimization on large scale sparse systems \cite{wu2020pyoptsparse}. Other solutions seek to reduce computational cost by reducing data dimensionality, reducing the dataset size and retaining the most relevant information. This size reduction allows for the use of highly efficient feature extraction algorithms which lower the computation cost even further, such as deep learning models which combine SVD with a neural network that can be used to extract the main features of a dataset in order to generate (predict) future snapshots at a much higher speed than standard numerical solvers \cite{abadia2022predictive,corrochano2023predictive}. A similar combination of SVD and deep learning can be used to enhance the spatial resolution of an experimental dataset to a desired size, which mainly coincides with the dimensions of its numerical simulation dataset \cite{diaz2023deep}. 

Another feature extraction algorithm that can be applied is principal component analysis (PCA), which is used in Ref. \cite{parente2013principal} to identify the most active
directions in multivariate datasets. PCA can also be used for filtering extracts dominant coherent structures to identify and fill in incorrect or missing measurements of experimental data, as demonstrated in Ref. \cite{scherl2020robust}. Other algorithms, such as dynamic mode decomposition (HODMD) \cite{LeClaincheVega17}, can be used by itself \cite{le2019new}, or combined with PCA to identify the main patterns that describe a flow \cite{corrochano2023higher}. This algorithm is also combined with neural networks in \cite{MataLeon2023,huang2022predictions}, in the same manner as SVD, to predict future data based on flow patterns. There are also other feature extraction algorithms that do not involve modal decomposition techniques, such as autoencoder models, which are used compress and reconstruct datasets, as demonstrated in Ref. \cite{munoz2023extraction,ae_modal}.

Current trends, which are vastly influenced by the evolution of the field of data science and its multiple industrial applications, show that experimental data can be measured by using a series of sensors which are used to collect information about the dynamic system object of study. Sensors can be use to measure the pressure distribution on the surface of a wind turbine blade \cite{soto2020determination}, to study a combustion process using particle image velocimetry (PIV) \cite{fang2018particle}, to measure turbulence in a turbulent boundary layer for aircraft design optimization \cite{woodward2023data}, to measure the drag forces of an airplane wing model \cite{botez2018morphing}, or to detect cavitation in a pump \cite{siano2017diagnostic}.   Reconstructed datasets created with data measured by sensors generally offer highly detailed temporal evolution data but, on the other hand, offer limited spatial information due to the spatial sparsity of the collected data.  Despite this sparsity, sometimes it may be interesting to exploit the spatial information contained in reconstructed datasets, since in many occasions this information can contain key data which is essential to solve specific problems, mainly related to data reconstruction. Some of the algorithms capable of completely reconstructing an experimental dataset with high precision just by using the key information contained in the sparse spatial data are gathered in ModelFLOWs-app \cite{hetherington2023modelflows}. This open-source sofwtare uses modal decomposition algorithms, or combines these with neural networks to reconstruct datasets, filling in missing data that sensors are not able to measure due to their range or due to a fault, by using the gappy SVD algorithm, or enhancing an experimental dataset resolution using the superresolution algorithm. Other solutions include the use of extreme machine learning (EML) autoencoders \cite{zhou2015compressed} to reconstruct data measured from sparse sensors, or shallow
neural network-based learning for data reconstruction with a limited amount of sensors \cite{erichson2019shallow}. 

To maximize the information collected from the dynamic system that is being studied, and even more to ensure that key information is being measured, strategic sensor placement is fundamental. In this article, the \textit{pysensors} Python library \cite{pysensors} is used to obtain an initial solution for optimal sensor placement. This library offers different sensors placement solution depending on the purpose, which can be data reconstruction or classification, taking into account the expense of some of the commonly used sensors, like sensors which are required to have high precision and accuracy \cite{arciniega2012deciphering}, or those which must be made with specific materials making them resistant to extreme environments \cite{wong2016integrated}, to name a few examples.

Being able to work with datasets that have been reduced in size is especially beneficial when applying different data analysis algorithms. On one hand, this offers the possibility of drastically reducing computation time and memory, which allows users to serialize algorithms instead of using parallelization techniques, sometimes complex to implement, or also allow the users using standard computers to solve problems. On the other hand, more data can be stored, which is helpful when retrieving the most relevant information from the analyzed system. For example, by augmenting the amount of data used to resolve a time series problem, more precise conclusions can be extracted regarding the behaviour of said time series, minimizing uncertainty errors which are related to the data itself. It also provides deeper knowledge about stochastic series, or data with noisy or chaotic components.

This article introduces a new method called low-cost SVD (lcSVD) that allows users to perform SVD with a very low computational cost. The method uses data collected by sparse sensors, which can be optimally, randomly, or equidistantly placed over the data, and then applies the lcSVD algorithm, minimizing computational cost, and reconstructs the final solution. This article presents a highly efficient and fast algorithm which calculates the optimal sensor positions that offer minimum reconstruction error, optimal sensor lcSVD (OS-lcSVD) that, based on a given number of sensors and a reconstruction error tolerance, is capable of finding the optimal number of sensors to be used and their optimal positioning, using minimum computational resources. 

Uncertainty quantification is used to analyze the reconstruction error based on the error data density, and to find the optimal number of sensors needed to reconstruct a given dataset with minimum error by implementing the elbow method in a similar way to how it is commonly used in the k-means clustering unsupervised machine learning algorithm to find the optimal number of clusters \cite{umargono2019k}.

This article is organized in the following way: the lcSVD methodology and its implementation to validate the optimal placement of sensors is detailed in Sec. \ref{sec:methodology}. The different datasets used to validate the developed method robustness, effectiveness and precision are described in Sec. \ref{sec:database}, followed up by Sec. \ref{sec:results}, where the results of this new method are displayed. These results include the optimal sensor number and placement, dataset reconstruction results, as well as computational cost metrics. Finally, the conclusions are presented in Sec. \ref{sec:conclusions}.

\section{Optimal sensor placement based on low-cost singular value decomposition \label{sec:methodology}}

Optimal sensor placement based on low-cost singular value decomposition (OS-lcSVD) is a new method to identify the optimal points to select in a dataset to reduce its spatial dimension, maximizing the information containing into such dataset. The original dataset can later be easily reconstructed. 

\begin{figure}[H]
    \centering
    \includegraphics[width=1\textwidth, angle=0]{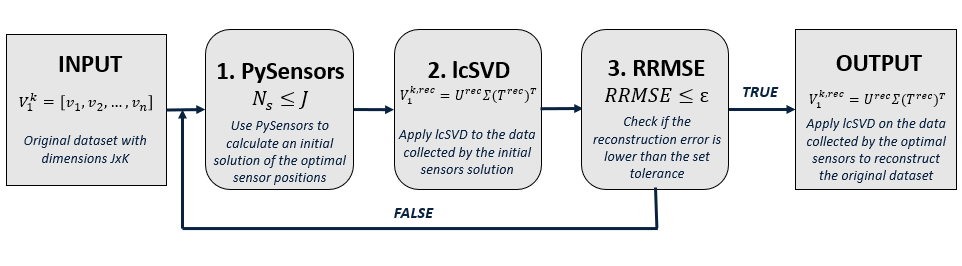}
    \caption{Optimal sensor placement based on low-cost SVD: sketch summarizing the methodology.}
    \label{fig:SKETCH}
\end{figure}

The applications of this algorithm is two-fold. On the one hand, it is possible to efficiently reducing data dimensionality, very useful in the case of numerical simulations, generally formed by a large number of degrees of freedom. Identifying the relevant points in the field, it is possible to reduce computational cost in numerical simulations related to time and memory storage, since only relevant points of the domain can be saved. Then, this information can be used to reconstruct the entire flow field when needed. On the other hand, sparse measurement are very common in experimental measurements when a bunch of sensors are placed along the region of interest. However, finding the optimal position where this sensors can be located to extract as much relevant information as possible it not an easy task.  The present algorithm provides a good approach capable to provide information about the optimal position of sensors to take experimental measurements, with the main goal of using later this information to reconstruct the complete two- or even three-dimensional flow field. Figure \ref{fig:SKETCH} shows an sketch summarizing the methodology.

\subsection{Data organization \label{sec:dataOrganization}}

A group of $K$ time-varying samples, known as snapshots, each defined as 

The data set collected, formed by $K$ temporal samples, are organized in matrix form as
\begin{equation}
    \bV_1^{K} = [\bv_{1}, \bv_{2}, \dots, \bv_{k}, \bv_{k+1}, \dots, \bv_{K-1}, \bv_{K}],
    \label{eq:SnapMatrix}
\end{equation}
where $\bV_1^{K}$ is known as the snapshot matrix and  each one of the $K$ temporal samples $\bv_{k}$ is known as snapshot. Each snapshot is formed by a vector with dimension $J=N_x \times N_y $ for two-dimensional problems and $J=N_{comp} \times N_x \times N_y \times N_z$ for three-dimensional cases, where $N_{comp}$ correspond to the number of components of the dataset (i.e., in a dataset formed by the pressure vector and the streamwise and normal velocity components $N_{comp}=3$), and $N_x$, $N_y$ and $N_z$ correspond to the spatial grid points distributed along the streamwise, normal and spanwise direction. When the dataset is formed by several components each one of these components is concatenated in columns.   

It is possible to generate a modified reduced snapshot matrix $\bar{\bV}_1^{K}$ with dimension $\bar{J}\times \bar{K}$, where $\bar{J}<J$ and $\bar{K}<K$. This dimension reduction can be carried out in several ways. For instance, it could be possible to use a uniform downsampling of the dataset, where $1$ every $p$ points is retained. Another alternative, but only possible to reduce the spatial dimenison of the data, is to use non-uniform downsampling, based on sensor or sparse measurements randomly organised along the whole spatial domain. A novel approach to select the optimal position of these sensors is presented below in Sec. \ref{sec:pysensors}. For simplicity, eq. (\ref{eq:SnapMatrix}) is repeated below for the modified reduced snapshot matrix as 
\begin{equation}
    \bar{\bV}_1^{K} = [\bar{\bv}_{1}, \bar{\bv}_{2}, \dots, \bar{\bv}_{k}, \bar{\bv}_{k+1}, \dots, \bar{\bv}_{K-1}, \bar{\bv}_{K}],
    \label{eq:SnapMatrixRed}
\end{equation}
where $\bar{\bv}_{k}$ represents a modified reduced snapshot, $\bar{\bv}_{k} \in \mathbb{R}^{\bar{J}}$.

\subsection{Singular value decomposition (SVD) and proper orthogonal decomposition (POD)\label{sec:SVD}}

Singular value decomposition (SVD) \cite{Sirovich87} is a factorization technique suitable for low-rank approximations, which can also be used to remove data redundancies and filter noisy or spurious artefacts. The method is also used to identify the Proper Orthogonal Decomposition (POD) modes. POD is a mathematical approach that was first introduced by Lumley\cite{Lumley} to identify coherent structures from turbulent flows. Although POD and SVD are terms used interchangeably in the literature, SVD is only one of the two possible technique that can be used to calculate the POD modes and the most popular one, since the other technique is associated with a high computational cost \cite{LeClaincheetalJFM20}.

SVD decomposes a collection of spatio-temporal data $\bv(x,y,z,t)$ into a set of proper orthogonal spatial modes, commonly referred to as SVD or POD modes. These modes are multiplied by the corresponding temporal coefficients to weight their temporal behavior as
\begin{equation}
\bv(x,y,z,t)\simeq \sum_{n=1}^N \bc_n(t)\bPhi_n(x,y,z),\label{eq:SnapMatrix0pod}
\end{equation}
where $\bPhi_n(x,y,z)$ and $\bc_n(t)$ represent the SVD modes and associated temporal coefficients.

The snapshot matrix $\bV_1^k$, eq. (\ref{eq:SnapMatrix}), is factorized using SVD algorithm as 
\begin{equation}
\bV_1^{K}\simeq\bW\,\bSigma\,\bT^\top,\label{eq:svd}
\end{equation}
where $()^\top$ denotes the matrix transpose, $\bW$ and $\bT$ are the matrices containing in columns the spatial SVD (or POD) modes and corresponding temporal coefficients (the columns of these matrices are orthonormal), and $\bSigma$ is a diagonal matrix containing the  singular values $\sigma_1,\cdots,\sigma_{N}$, with $N\leq min(J,K)$ as the number of SVD modes retained, which are ranked (by the singular values) in decreasing order in the previous factorization. It is remarkable that $\bW^\top\bW = \bT^\top\bT=$ are unit matrices of dimension $N\times N$.

The number of $N$ SVD modes retained  is defined based on a tolerance $\varepsilon_{svd}$ as
\begin{equation}
\sigma_{N+1}/\sigma_{1}\leq \varepsilon_{svd}.\label{eq:TOLsvd}
\end{equation}
This tolerance will reduce the data dimensionality from $J$ to $N$, where $N$ is referred as the spatial complexity \cite{LeClaincheVega17}. In experimental databases, to clean the noise, the tolerance can be selected similar to the level of noise or uncertainty of the experiments (the SVD modes with smallest singular value, neglected, generally represent noise from the data set, while the main features are contained in the modes with largest singular value). In the analysis of turbulent flows, the tolerance could be related to the size of the coherent structures of the flow retained. It is remarkable that $N$ can also be determined using other criteria: see for instance these choices in Ref. \cite{PCA}, which are described in the context of Principal Component Analysis (PCA) algorithm (similar as SVD).

\subsection{Low-cost singular value decomposition (lcSVD)\label{sec:lowcostSVD}}

Low-cost SVD (lcSVD) is an extension of SVD that is suitable to analyse reduced data sets and to reconstruct their original solution, as presented in Ref. \cite{Lupod}. In that way, the computational cost of the algorithm SVD can be strongly diminished, and the method can be applied to analyse large databases using reduced computational resources.

The method applies SVD to the modified reduced snapshot matrix $\bar{\bV}_1^{K} \in \mathbb{R}^{\bar J\times \bar K}$ eq. (\ref{eq:SnapMatrixRed}), then uses the decomposed dataset to reconstruct the original snapshot matrix  $\bar{\bV}_1^{K} \in \mathbb{R}^{J\times K}$  eq. (\ref{eq:SnapMatrix}). 

It is possible to reduce only one of the dimensions of the snapshot matrix. This is called as semi-reduced snapshot matrix, and is written as $\bar{\bV}_{1}^{K,{J\bar K}}\in \mathbb{R}^{J\times \bar K}$ or $\bar{\bV}_{1}^{K,{\bar J K}}\in \mathbb{R}^{\bar J\times  K}$, for reductions in columns (space) or rows (time), respectively.

The algorithm is as follows.
\begin{itemize}
    \item {\em Step 1: SVD in the reduced dataset.} SVD is applied to the reduced dimension snapshot matrix eq. (\ref{eq:SnapMatrixRed}) as
    \begin{equation}
    \bar{\bV}_1^{K}\simeq\bar{\bW}\,\bar{\bSigma}\,\bar{\bT}^\top,\label{eq:redSVD}
    \end{equation}
    where $\bar{\bW}^\top\bar{\bW} = \bar{\bT}^\top\bar{\bT}=$ are unit matrices of dimension $\bar{N}\times \bar{N}$.  The number of SVD modes retained, $\bar{N}$, is calculated based on a tolerance as in eq. (\ref{eq:TOLsvd}).
    \item {\em Step 2: normalization of SVD modes.} Matrix $\bar{\bSigma}$ can be ill-conditioned when small singular values are retained. As consequence, the SVD modes calculated in $\bar{\bW}$ could be slightly non orthogonal due to round-off errors. QR factorization is then used to re-orthonormalize these modes as $ \bar{\bW}=\bQ^W \bR^W$, which leads to 
    \begin{equation}
        \bar{\bW}=\bar{\bW} (\bR_{\bar N}^W)^{-1},
    \end{equation}
    where $\bR_{\bar N}^W \in \mathbb{R}^{\bar N\times \bar K}$ (similar to SVD, only $\bar N$ modes are retained). 
    \item {\em Step 3: normalization of SVD temporal coefficients.} 
      Similarly to Step 2,  SVD temporal coefficients calculated in $\bar \bT$ could be slightly non orthogonal, so again, QR factorization is then used to re-orthonormalize these modes as  as $ \bar{\bT}=\bQ^T \bR^T$, which leads to 
    \begin{equation}
        \bar{\bT}=\bar{\bT} (\bR_{\bar N}^T)^{-1},
    \end{equation}
    where again $\bR_{\bar N}^T\in \mathbb{R}^{\bar N\times \bar K}$.
    Additionally, some differences could be found in the sign of these temporal coefficients depending on the type of calculations carried out that could affect to the next re-construction of the original dataset. To avoid this type of conflicts, an additional step can be carried out, where sings in  $\bar{\bT}$ are updated as:
     \begin{equation}
        \bar{\bT}=\bar{\bT} \text{ sign}(\text{diag} (\bar{\bSigma})), %
    \end{equation}
    where sign($\cdot$) and diag($\cdot$) correspond to the sign and diagonal of a matrix. To avoid conflicts and information losses with the calculations of these signs in $\bar{\bSigma})$, we recommend to implement $(\bar{\bW}^\top  \bar{\bV}_1^{K} \bar{\bT}$, rather than directly using $\bar{\bSigma}$, although this is dependant on the type of coding language used in this implementation). 
    \item {\em Step 4: recover SVD modes.} The SVD modes with enlarged spatial dimension (matrix $\bW$ from eq. (\ref{eq:svd})) are recovered as:
    \begin{equation}
        \bW\simeq \bW^{rec}= (\bar{\bV}_{1}^{K,{J\bar K}})^\top \bar{\bT} (\bar{\bSigma})^{-1},\label{eq:Wrec}
    \end{equation}
    where $\bW^{rec} \in \mathbb{R}^{J \times \bar N}$.
    
    \item {\em Step 5: recover temporal coefficients of SVD modes.} The temporal coefficients of SVD modes with enlarged spatial dimension (matrix $\bT$ from eq. (\ref{eq:svd})) are recovered as:
    \begin{equation}
        \bT\simeq=\bT^{rec}\simeq (\bar{\bV}_{1}^{K,{\bar J K}})^\top \bar{\bW} (\bar{\bSigma})^{-1},\label{eq:Trec}
    \end{equation}
    where $\bT^{rec} \in \mathbb{R}^{K \times \bar N}$.

    \item {\em Step 6: reconstruction of the original dataset.}
    Using the reconstructed SVD modes and temporal coefficients eqs. (\ref{eq:Wrec})-(\ref{eq:Trec}), as well as the reduced matrix containing the singular values, it is possible to reconstruct the original dataset $\bar{\bV}_{1}^{K}$ from eq. (\ref{eq:svd}) as
    \begin{equation}
\bV_1^{K}\simeq \bV_1^{K, rec}= \bW^{rec}\,\bar \bSigma\,(\bT^{rec})^\top.\label{eq:svdRec}
\end{equation}
    \item{\em Step 7: reconstruction error measurement.} Based on uncertainty quantification which gives a global view of the reconstruction error component by component. If the data is two-dimensional and has more than one velocity component, the snapshots are re-organized in a fourth-order $J_1\times J_2\times J_3\times K$-tensor $\bV$, whose components $\bV_{j_1j_2j_3k}$ are defined as
\begin{equation}
\bV_{1j_2j_3k}=v_x(x_{j_2},y_{j_3},t_k),\quad \bV_{2j_2j_3k}=v_y(x_{j_2},y_{j_3},t_k).\label{eq:velocities2D}
\end{equation}

In this case, indexes $j_1, j_2, j_3$ and $k$ are the labels of the velocity components ($j_1=1,2$, where $J_1=2$), the discrete values of the spatial coordinates, and the values of time. Three-dimensional databases are organized in a fifth-order $J_1\times J_2\times J_3\times J_4\times K$-tensor $\bV$,
 whose components $\bV_{j_1j_2j_3j_4k}$ are defined as 
 \beqn
 \begin{split}
\bV_{1j_2j_3k}&=v_1(x_{j_2},y_{j_3},z_{j_4},t_k),\\ \bV_{2j_2j_3k}&=v_2(x_{j_2},y_{j_3},z_{j_4},t_k),\\
\cdots\\
\bV_{j_1j_2j_3k}&=v_{j_1}(x_{j_2},y_{j_3},z_{j_4},t_k),\\
\cdots\\
\bV_{J_1j_2j_3k}&=v_{J_1}(x_{j_2},y_{j_3},z_{j_4},t_k).\label{eq:velocity3D}
\end{split}
\eeqn
$J_1$ is the number of velocity components, the indexes $j_2$, $j_3$ and $j_4$ correspond to the discrete values of the three spatial coordinates, $x$, $y$ and $z$, and $k$ is the index representing the time instant.

Using the two-dimensional case, the reconstruction errors for the streamwise and normal velocity, $\epsilon_{u}$ and $\epsilon_{v}$, are calculated as the difference between the original and reconstructed data are calculated as
\begin{equation}
\epsilon_{u} = \bV_{1j_2j_3k} - \bV^{rec}_{1j_2j_3k}, \quad \epsilon_{v} = \bV_{2j_2j_3k} - \bV^{rec}_{2j_2j_3k},\label{eq:epsUV} 
\end{equation}

and are then normalized based on the free flowing fluid velocity $U_{\inf}$

\begin{equation}
\frac{\epsilon_{u}}{U_{\inf}}, \quad \frac{\epsilon_{v}}{U_{\inf}}.\label{eq:epsnormUV} 
\end{equation}

The probability density curves of the reconstruction errors of each component of the reconstructed dataset (eq. \eqref{eq:epsnormUV}) are then plotted. These errors tend to follow a normal distribution. A narrow and tall curve centered in 0 means that there is a higher probability of obtaining a low reconstruction error, and vice-versa. 

Error contour maps are also graphed to visualize the point by point reconstruction absolute error for each snapshot, which is computed as so:

\begin{equation}
    Absolute \; Error_{j_1j_2j_3k}= | \bV_{j_1j_2j_3k}-\bV_{j_1j_2j_3k}^{rec} |,
    \label{eq:pointrrmse}
\end{equation}
where, as previously explained, $j_1$, $j_2$ and $j_3$ are the indexes for the velocity component, $x$ and $y$ coordinates, respectively, while $k$ is used to index the snapshot. The $|$ operand represents the absolute value.

\end{itemize}

\subsection{Optimal sensor placement \label{sec:pysensors}}

This section introduces an algorithm to identify the optimal regions in a field to locate a bunch of sensors to represent a signal, with the idea of using this information to later reconstruct the entire field. The algorithm presented here is based on a dimensionality reduction using SVD, and is combined with QR algorithm. This algorithm is part of the open-source code {\em pysensors} \cite{pysensors}. More details about this code and the different methods that can be used for optimal sensor placement can be found in Ref. \cite{pysensorsMethod}.

Considering a signal varying in time $\bv_k$, organized in the snapshot matrix $\bV_1^k$ eq. (\ref{eq:SnapMatrix}), it is possible to collect a set  of sparse measurements $\by \in \mathbb{R}^{P \times K}$, given by a measurement matrix $\bC \in \mathbb{R}^{P \times J}$, where $P$ is the number of sensors, which determines the sensor location, as
\begin{equation}
 \by=\bC \bV_1^K. \label{eq:sens}   
\end{equation}
For $P<K$ the system is undetermined and the solutions are infinite. Applying SVD eq. (\ref{eq:svd}) into the previous equation, leads to
\begin{equation}
   \by=\bC \bV_1^K\simeq \bC \bW\,\bSigma\,\bT^\top=\bC \bW \hat\bT, \label{eq:sensSVD}    
\end{equation}
where, as mentioned before, only $N$ SVD modes are retained, thus $\bW^\top\bW = \bT^\top\bT=$ are unit matrices of dimension $N\times N$. 

Solving an optimization problem, it is possible to determine the measurement matrix $\bC$. In the present article, we use a method based on the QR factorization with column pivoting (see more applications of QR to solve least-squares problems and optimal sensor placement in Refs. \cite{R1pys,R2pys,R3pys,R4pys}) applied to the SVD (or POD) basis of modes. More specifically, the algorithm presented here (from {\em pysensors})  is an extension of Q-DEIM algorithm that is capable to solve (in addition to the standard case where $P\leq N$), cases where the number of sensors exceeds the number of SVD modes, $P>N$.

The QR factorization with column pivoting provides $N$ pivots, which correspond to the point sensors, that best sample the $N$ SVD modes of the SVD basis $\bW$ as
\begin{equation}
    \bW^\top \bC^\top= \bQ \bR, 
\end{equation}
where $\bQ$, $\bR$ and $\bC$ correspond to a unitary matrix, an upper triangular matrix and a column permutation matrix, respectively. 

Similarly, For the case of oversampled matrix, $P>N$, the the QR method is then applied as 
\begin{equation}
    (\bW \bW^\top) \bC^\top= \bQ \bR.
\end{equation}

Given a certain number of sensors $P$ (selected by the user), randomly located, it is possible to obtain multiple optimal solutions  (each one associated to each one of the initial conditions where sensors are placed). The present article combines this optimal sensor placement algorithm with the lcSVD method to select the optimal sensor placement that minimizes the reconstruction of the original matrix, as explained below.  

\subsection{Optimal sensor placement based on low-cost singular value decomposition (OS-lcSVD): the algorithm\label{sec:OptimSensSVDalgorithm}}

The novel methodology proposed in this article for optimal sensor placement, namely OS-lcSVD method, combines {\em PySensors}, a python package for spare sensor placement \cite{pysensors,pysensorsMethod}, and a low-cost singular value decomposition (lc-SVD) method, capable to perform SVD in a reduced dimension dataset, and using this information to reconstruct the original flow field. 

The algorithm is described as follows.
\begin{itemize}
    \item {\em Step 1: select optimal sensors.} The user selects the number of sensors to reduce the data dimension, $N_s = \bar J$ ($J$ is the maximum possible number of sensors). The algorithm {\em PySensors} is then used to select the optimal position of these sensors based on a certain initial condition (selected randomly) of the position of the sensors over the entire domain. The modified reduced snapshot matrix eq. (\ref{eq:SnapMatrixRed}), $\bar \bV_1^{K}\in \mathbb{R}^{\bar J\times \bar K}$, is then obtained at this step.
    \item {\em Step 2: apply lcSVD.} The method lcSVD is then applied over the previous reduced matrix. The output is the reconstructed original snapshot matrix, as presented eq. (\ref{eq:svdRec}), $\bV_1^{K,rec}\in \mathbb{R}^{J\times K}$.
    \item {\em Step 3: Compute the error.} Root mean square error (RMSE) is then computed, comparing the original and reconstructed snapshot matrix, as
    \begin{equation}
        RRMSE=\frac{\|\bV_1^{K}-\bV_1^{K,rec}\|_2}{\| \bV_1^{K}\|_2},
        \label{eq:rrmse}
    \end{equation}
    where $\| \cdot \|_2$ corresponds to the L2-norm. If $RRMSE<\varepsilon$, where $\varepsilon$ is a tolerance given by the user, the method considers that the optimal sensors positions have been calculated. Otherwise, the method returns back to {\em Step 1} with a new initial position for the sensors (randomly selected).
\end{itemize}

Using lcSVD as part of this algorithm is a key point. On the one hand, it allows very fast implementations of SVD, and on the other hand, provides optimal reconstruction of the original solution.

\subsection{Optimal number of sensors estimation \label{sec:elbow}}

The optimal number of sensors needed to reconstruct a certain dataset can be estimated in two ways: using a reconstruction error tolerance, or using uncertainty quantification data.

In the first case, the user will set a reconstruction error tolerance and OS-lcSVD will calculate the optimal number of sensors, being the minimum number of optimally positioned sensors which give a reconstruction error, calculated using eq. \eqref{eq:rrmse}, inferior to the input tolerance $\varepsilon$. This technique can be used on any type of dataset.

For more simple datasets, specifically those with little noise, uncertainty evaluation can be used to estimate the optimal number of sensors needed to reconstruct a dataset using the elbow method technique, as explained below in a similar manner as implemented to find the optimal number of clusters in k-means clustering, an unsupervised machine learning algorithm that groups together similar data forming clusters (see Ref. \cite{kuraria2018centroid}). 

Given a range of sensors $s = 1, 2, 3, ... S$ provided by the user, and a number of SVD modes proportional to the number of sensors, the steps described in Sec. \ref{sec:OptimSensSVDalgorithm}, followed by those in Sec. \ref{sec:lowcostSVD}, are applied to the input data, which is reconstructed for each case of number of sensors $s$. Next, the reconstruction error $\epsilon$ as described in Sec. \ref{sec:lowcostSVD}, step 7, eq. (\ref{eq:epsUV}). The bias is then is calculated for each case of number of sensors, as

\begin{equation}
   \overline{\epsilon_{u}^{s}} = E[\epsilon_{u}^{s}], \quad \overline{\epsilon_{v}^{s}} = E[\epsilon_{v}^{s}],
   \label{eq:bias}    
\end{equation}

where $\epsilon_{u}^{s}$ and $\epsilon_{v}^{s}$ are the streamwise and normal velocity reconstruction errors, and $\overline{\epsilon_{u}^{s}}$ and $\overline{\epsilon_{v}^{s}}$ are the bias values for each velocity component reconstruction error for each case of number of sensors $s$. The $E$ operand symbolizes the long-term average or mean. The uncertainty is calculated as

\begin{equation}
   \sigma_{u}^{s} = \sqrt{E[(\epsilon_{u}^{s} - \overline{\epsilon_{u}^{s}})^{2}]}, \quad 
   \sigma_{v}^{s} = \sqrt{E[(\epsilon_{v}^{s} - \overline{\epsilon_{u}^{s}})^{2}]},
   \label{eq:std}    
\end{equation}

Where $\sigma_{u}^{s}$ and $\sigma_{v}^{s}$ correspond to the streamwise and normal velocity reconstruction error uncertainty values for each number of sensors $s$, respectively. When these uncertainty values are plotted an elbow can be visualized, where the inflection point allows the user to estimate the optimal number of sensors to use for the minimum reconstruction error. As mentioned previously, this technique works with simple datasets, precisely those will little noise, since the reconstruction error uncertainty is initially low and will stagnate sooner for a smaller amount of sensors, creating the elbow.




\section{Test cases\label{sec:database}}
To show the accuracy, robustness, and low computational cost (compared to standard SVD) of the presented method, diverse datasets have been used for this purpose. These datasets consider experimental data and numerical simulations, laminar and turbulent flows, and two- and three-dimensional databases. In particular, the test cases studied are two circular cylinder numerical datasets, consisting of a two-dimensional Re $= 100$ cylinder and a three-dimensional Re $= 280$ cylinder, an experimental turbulent boundary layer, a turbulent jet large eddy simulation, and two experimental turbulent cylinders, with Re $= 4000$ and Re $= 2600$.

All test cases consist in fluid dynamics experiments or numerical simulations, and are governed by the Navier-Stokes equations. These equations for a viscous, incompressible and Newtonian flow are:

\begin{equation}
\vec{\bV} \cdot \nabla = 0,
\end{equation}

\begin{equation}
\frac{\partial u}{\partial t} + (\vec{\bV} \cdot \nabla) \vec{\bV} = -\nabla p + \frac{1}{Re} \Delta \vec{\bV},
\end{equation}

where $\vec{\bV}$ is the velocity vector, $p$ is the pressure, and Re is the Reynolds number, which is defined differently for each case. These equations are nondimensionalized using the characteristic length $L$ and time $L/U$, where $U$ is the characteristic or free stream velocity for each case.

Since one of the test cases consists of a turbulent boundary layer, a third equation must be added to the previously defined equations. This equation is the energy conservation equation, which is defined as so:

\begin{equation}
\rho c_p \left(\frac{\partial T}{\partial t} + (\vec{\bV} \cdot \nabla)T\right) = k \nabla^2 T + \phi,
\end{equation}

where $\rho$ and $T$ are the density and temperature of the fluid, respectively, $c_p$ is the specific heat capacity at constant pressure, $k$ is the
thermal conductivity coefficient, and $\phi$ represents the dissipation term, which accounts for the conversion of kinetic energy into thermal energy due to the viscous effects within the fluid.

\subsection{Circular cylinder datasets \label{Cyl2D3D}}
The first datasets, which consist of a numerical database solving both a two- and three-dimensional flow passing a circular cylinder, is extracted from Ref. \cite{VegaLeClaincheBook20}. This test case has been selected because is generally used as a benchmark problem to validate methodologies. The dynamics of the cylinder are tightly related to the concept of the Reynolds number, defined with the cylinder diameter $D$. The flow is steady for low Reynolds numbers. From Re $ \approx 46$, a Hofp bifurcation that produces an unsteady flow, which is conducted by a von Karman vortex street \cite{jackson1987finite}, emerges. These oscillations remain two-dimensional until Re $ \approx 189$ where a second bifurcation occurs and the flow develops into a three-dimensional flow for some specific wavelengths in the spanwise direction \cite{barkley1996three}. 

The flow becomes fully three-dimensional for some specific wave numbers at Re $>180$, and fully turbulent for high Reynolds values. In this research two numerical simulations have been analysed, the two-dimensional cylinder at Re $ = 100$, and the three-dimensional cylinder at Re $ = 280$. In both of these cases, numerical simulations have been carried out  using the open-source solver Nek5000 \cite{Nek5000} to solve the incompressible Navier-Stokes equations which define the behaviour of the flow. This solver uses spectral elements methods as spatial discretization.

The boundary conditions configured in the simulation for the cylinder surface are Dirichlet for velocity ($u = v = w = 0$) and Neumann for pressure. The conditions in the inlet, upper and lower boundaries of the domain are the same: $u = 1$, $v = w = 0$ for the streamwise, normal and spanwise velocities, respectively, and Neumann condition for pressure. The conditions in the outlet are Dirichlet for pressure and Neumann for velocity. The domain of the computational simulations is composed by 600 rectangular elements, each one of these is discretized using the polynomial order $\Pi = 9$. The dimensions of the computational domain are non-dimensionalized with the diameter of the cylinder. The size of the domain in the normal direction is constant $L_{y} = 15D$, and extends in the streamwise direction from $L_{x} = 15D$ upstream of the cylinder to $L_{x} = 50D$ downstream.

\begin{figure}[H]
    \centering
    \includegraphics[width=1\textwidth, angle=0]{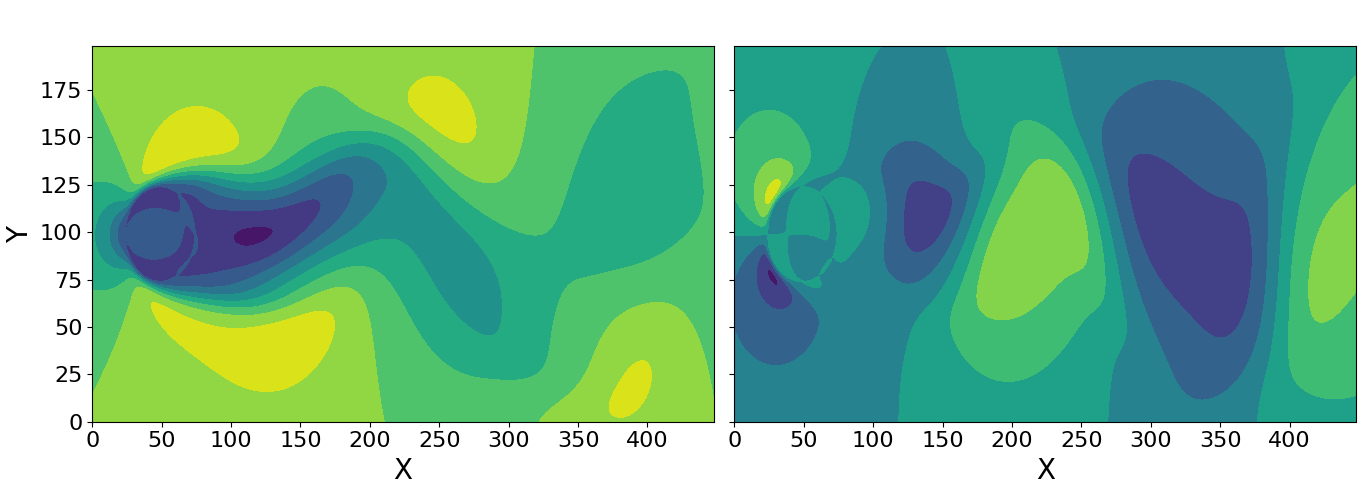}
    \caption{Streamwise (left) and normal (right) velocities of a representative snapshot of the two-dimensional Re $ = 100$ cylinder dataset from Ref. \cite{VegaLeClaincheBook20}.}
    \label{fig:cyl2ddataset}
\end{figure}

As mentioned, the two-dimensional cylinder dataset represents the saturated flow around a circular cylinder with Reynolds number Re $ = 100$. The dataset analysed is composed by $N_{t} = 150$ snapshots equidistant in time snapshots with an interval of $\Delta t = 0.2$. The  dimensions are $N_{x} = 449$ points in the streamwise direction and $N_{y} = 199$ points in the normal direction. The two variables ($N_{comp} = 2$) used in the reconstruction of the dataset correspond to the streamwise velocity $U$, and normal velocity $V$. Figure \ref{fig:cyl2ddataset} shows a representative snapshot of this database.

\begin{figure}[H]
    \centering
    \includegraphics[width=1\textwidth, angle=0]{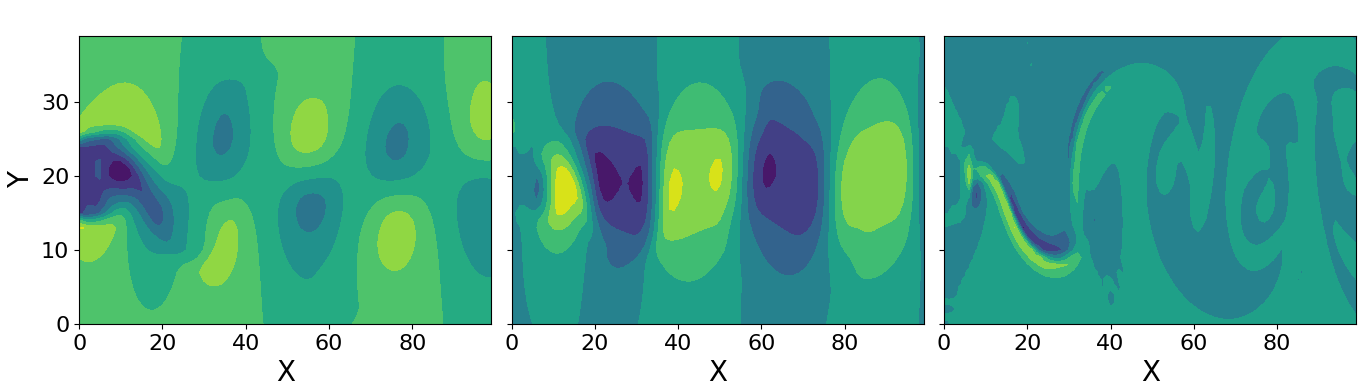}
    \caption{Streamwise (left), normal (middle) and spanwise (right) velocities of a representative snapshot of the three-dimensional Re $ = 280$ cylinder dataset from Ref. \cite{VegaLeClaincheBook20} in the $XY$ plane at $z = 32$.}
    \label{fig:cyl3ddataset}
\end{figure}

The three-dimensional cylinder dataset represents the three-dimensional flow past a circular cylinder at Re $ = 280$. The information that has been selected for this test case is the saturated regime of the numerical simulation, which is gathered in the last $299$ snapshots, equidistant in time with step size $\Delta t = 1$. The dimension of the three-dimensional cylinder dataset are $N_{x} = 100$ points in the streamwise direction, $N_{y} = 40$ in the normal direction, and $N_{z} = 64$ in the spanwise direction. The flow field velocity components are defined by $U$ for the streamwise velocity, $V$ for the normal velocity, and $W$ for the spanwise velocity ($N_{comp} = 3$). Figure  \ref{fig:cyl3ddataset} shows a representative snapshot of this database. 

\subsection{Turbulent boundary layer using planar particle image velocimetry\label{BL_PIV}}

The second dataset is an experimental turbulent boundary layer that has been extracted from Ref. \cite{towne2023database}, where details of the experiments can be found. 
To generate this dataset, a turbulent boundary layer was artificially transitioned, and then developed over $2.5 m$ on a flat plate located in the center of the wind tunnel section before making it to the PIV field of view. The boundary layer was developed in a nominally zero-pressure-gradient environment \cite{rodriguez2020development}. Time-resolved planar PIV (TR-PIV) was used to measure the velocity field in the streamwise–wall-normal.

\begin{figure}[H]
    \centering
    \includegraphics[width=1\textwidth, angle=0]{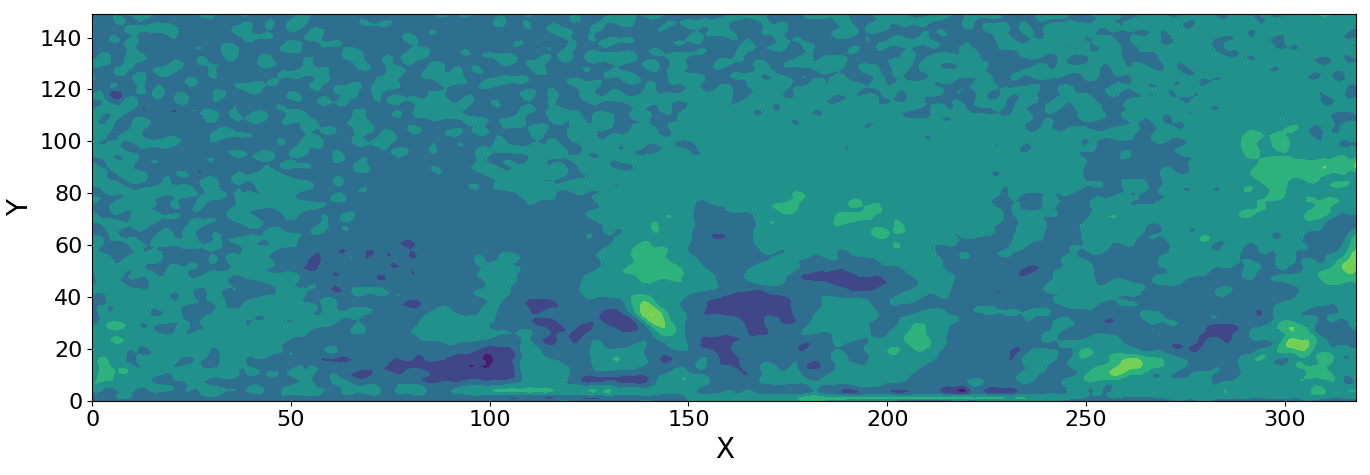}
    \caption{Streamwise velocity of a representative snapshot of the turbulent boundary layer dataset from Ref. \cite{towne2023database}.}
    \label{fig:bldataset}
\end{figure}

This dataset consists of $N_{t} = 6000$ snapshots formed by the streamwise  velocity field ($N_{comp} = 1$). The time step between snapshots is $\Delta t = 0.44$.
Each snapshot consists of $N_{x} = 319$ points along the streamwise direction and $N_{y} = 150$ points along the wall-normal direction. The flow parameters for this case are: the 99\% boundary layer thickness $\delta = 49 mm$, the friction velocity $u_{\tau} = 0.1827 m/s$, the free stream velocity $U_{\infty} = 3.71 m/s$, and the friction Reynolds number $Re_{\tau} = \delta u_{\tau} / \nu \approx  605$, where $\nu$ is the kinematic viscosity. The $u_{\tau}$ was computed using the Clauser method. Figure \ref{fig:bldataset} shows a representative snapshot of the dataset.


\subsection{Turbulent jet large eddy simulation\label{JetLES}}
This next dataset is a numerical database showing an isothermal subsonic jets issued from a round nozzle of exit diameter $D$ = 50 $mm$ and is extracted from Ref. \cite{bres2018importance}, where more details can be found. 

This simulation was conducted using the large eddy simulation (LES) compressible flow solver \emph{CharLES}, which solves the spatially-filtered compressible Navier-Stokes equations on unstructured grids using a finite-volume method and third-order Runge-Kutta time integration. The simulation used approximately sixteen million control volumes and was run for a duration of $2000$ acoustic time units ($tc_{\infty}/D$).

\begin{figure}[H]
    \centering
    \includegraphics[width=1\textwidth, angle=0]{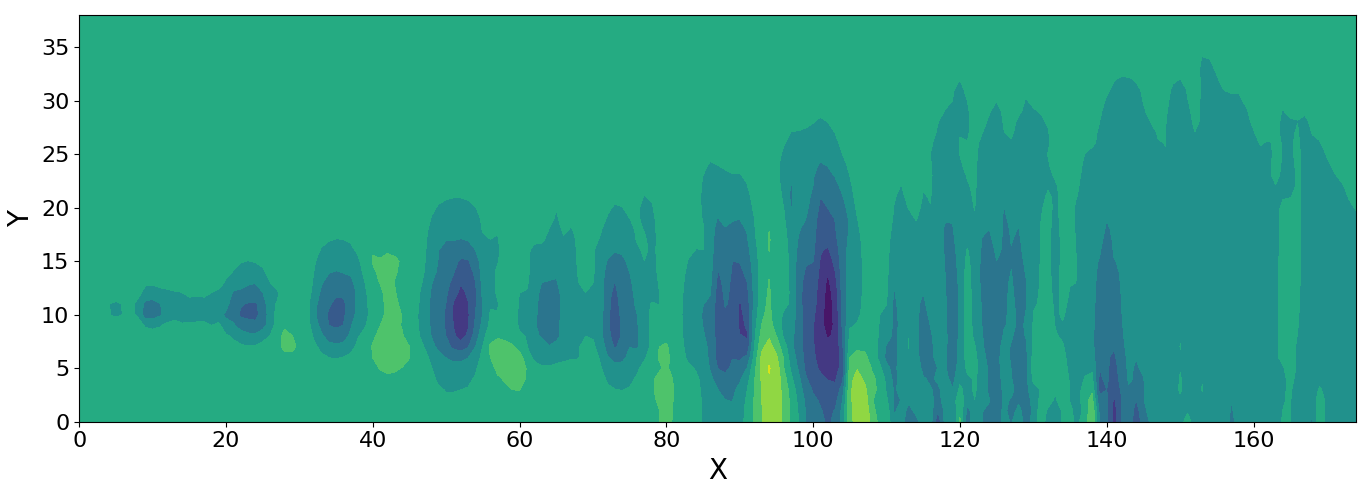}
    \caption{Streamwise velocity of a representative snapshot of the turbulent jet large eddy simulation dataset from Ref. \cite{bres2018importance}}
    \label{fig:jetdataset}
\end{figure}

This LES dataset consists of $5000$ snapshots of the jet's streamwise velocity ($N_{comp} = 1$) sampled every $\Delta t = 0.2$ acoustic time units on a structured cylindrical output grid that approximately mimics the underlying LES resolution and extends a distance of $30D$ in the streamwise direction and $6D$ in the radial direction, respectively. The simulation settings were set to match the experimental operational conditions, which are defined in terms of the nozzle pressure ratio, NPR = $P_{t}/P_{\infty} = 1.7$, and nozzle temperature ratio, NTR = $T_{t}/T_{\infty} = 1.15$, where the subscripts $t$ and $\infty$ refer to the stagnation (total) and free-stream (ambient) conditions, respectively. The jet is isothermal ($T/T_{\infty} = 1.0$), and the jet Mach number is $M$ = $U/c = 0.9$, where $U$ is the mean (time-averaged) jet exit streamwise velocity and $c$ is the speed of sound. With these conditions, the Reynolds number is Re $ = \rho U D/\mu \approx 10^{6}$. Each one of the  $N_{t} = 5000$ snapshots contains $N_{x} = 175$ points in the horizontal plane and $N_{y} = 39$ points in the vertical plane. Figure \ref{fig:jetdataset} shows a representative snapshot of the flow field.

The round nozzle geometry, with output centred at ($x$, $r$) = (0, 0), is explicitly included in the axisymmetric computational domain, which stretches from $-10D$ to $50D$ in the streamwise ($x$) direction and expands from $20D$ to $40D$ in the radial direction. A very slow co-flow at Mach number $M_{\infty} = 0.009$ is imposed outside the nozzle in the simulation to prevent spurious re-circulation and facilitate flow entrainment.

The Vreman sub-grid model \cite{vreman2004} is used to account for the physical effects of unresolved turbulence on the resolved flow, with constant coefficient set to $c = 0.07$. A constant turbulent Prandtl number $Pr = c_p \mu / k = 0.9$ is selected to complete the energy equation. To avoid spurious reflections at the downstream boundary of the computational domain, a damping function \cite{freund1997, mani2012} is applied in the outflow buffer zone as a source term in the governing equations. In addition, the numerical operators are switched to lower-order dissipative discretization in the sponge zone for $x/D > 31$ and $r/D > 7$, to further damp turbulent structures and sound waves.

All solid surfaces are regarded as adiabatic no-slip walls. In the preliminary parametric study, the far-field noise at $50D$ from the nozzle output was calculated for three different FW-H surfaces which consisted of a cylindrical surface of radius $0.65D$ up $x/D = 0$, continued by a conical surface that extended to $x/D = 30$ with different spreading rates of 0.11, 0.14 and 0.17. The jet spreading rate estimates are used as a basis to select the slopes \cite{zaman1998,zaman1999}.

\subsection{Transient turbulent circular cylinder \label{VKIdatasets}}
The final dataset is presented in Ref. \cite{mendez2020multiscale}, and consists in an experimental database representing the turbulent flow passing a circular cylinder of $D = 5 mm$ diameter and $L = 20 cm$ length in transient conditions, with a differing free stream velocity.

The experiment was carried out in the low speed wind tunnel of the Von Karman Institute. The dataset consists of the study of a turbulent flow passing a two-dimensional cylinder at two different Reynolds numbers, as well as the transient regime between both of these. These are Re $ \approx 4000$ and Re $ = 2600$, where the vortex shedding frequency goes from 450 $Hz$ to 303 $Hz$, corresponding to a Strouhal number of around $St = fd/U_{\infty} \approx 0.19$ in both regimes, respectively. The test's full Reynolds number range falls inside the region of three-dimensional vortex shedding \cite{williamson1996vortex}. In this test case, the free stream velocity $U_{\infty}$ transitions between two steady state conditions, precisely from $U_{\infty} = 12.1 \pm 3$\% to $U_{\infty} = 7.9 \pm 3$ \% $m/s$, with the transition between steady states being a smooth step of approximately 1 second.

\begin{figure}[H]
    \centering
    \includegraphics[width=1\textwidth, angle=0]{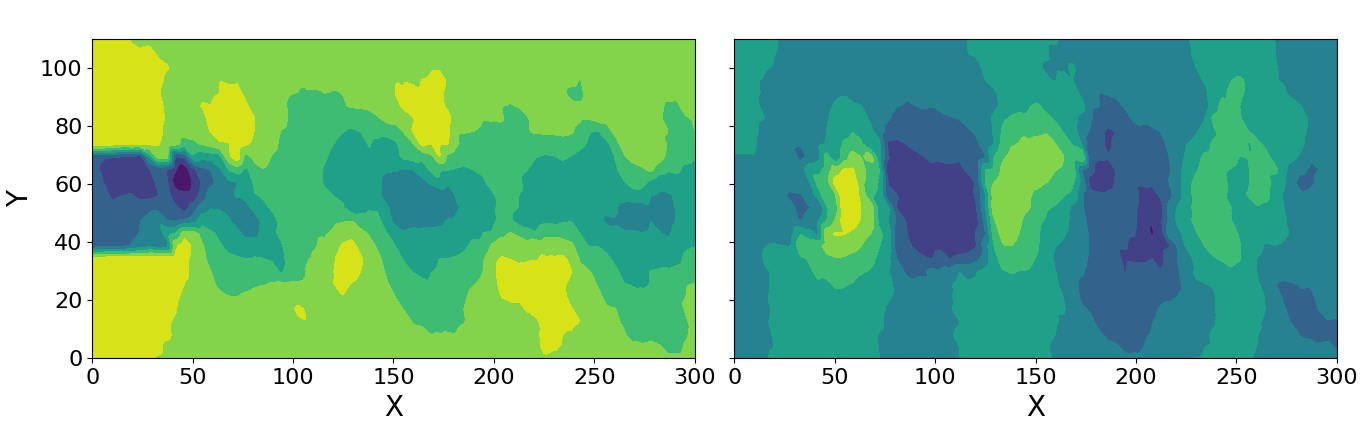}
    \caption{Streamwise (left) and normal (right) velocities of a representative snapshot of the turbulent Re $ = 4000$ cylinder dataset from Ref. \cite{mendez2020multiscale} (state one).}
    \label{fig:vki4000dataset}
\end{figure}

\begin{figure}[H]
    \centering
    \includegraphics[width=1\textwidth, angle=0]{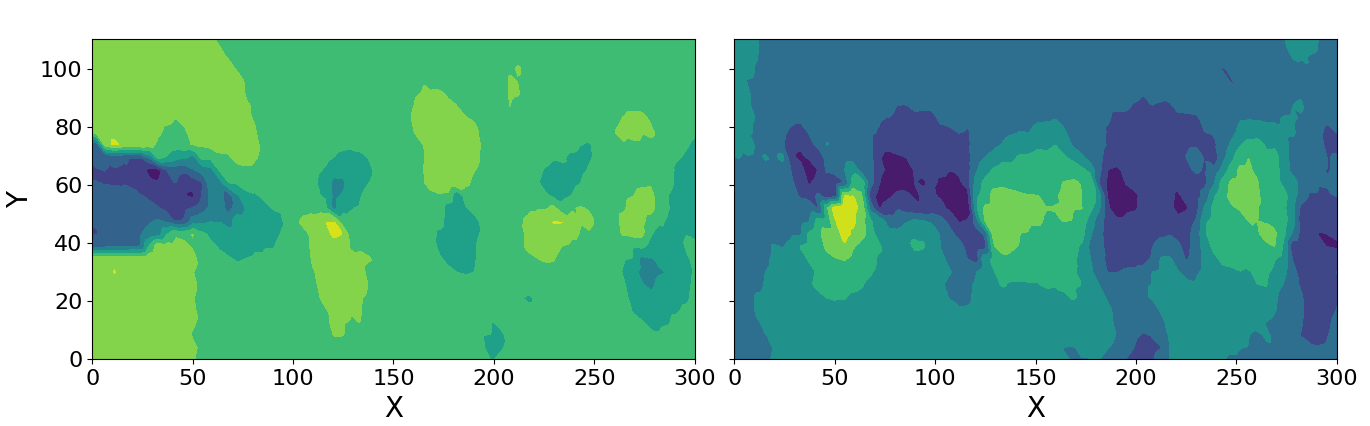}
    \caption{Streamwise (left) and normal (right) velocities of a representative snapshot of the turbulent Re $ = 2600$ cylinder dataset from Ref. \cite{mendez2020multiscale} (state two).}
    \label{fig:vki2600dataset}
\end{figure}

This dataset has been split into three parts: steady state one, where Re $ \approx 4000$ (Fig. \ref{fig:vki4000dataset}), transient state, and steady state two, with Re $ = 2600$ (Fig. \ref{fig:vki2600dataset}). To demonstrate the effectiveness of the method presented in this article, only steady states one and two have been used.

The dimension of the experimental domain are $N_{x}$ = 301 points in the streamwise direction and $N_{y}$ = 111 in the normal direction. This dataset consists of two velocity components ($N_{comp} = 2$), the streamwise velocity $U$ and normal velocity $V$, which were measured during 4000 snapshots for the steady state one dataset and 5200 snapshots for steady state two. Each snapshot is taken with time $\Delta t = 0.33$.

\section{Results\label{sec:results}}

This section presents the results of lcSVD when applied to each one of the test cases. Starting off with the optimal sensor positions calculated by applying OS-lcSVD. Next, the optimal number of sensor is obtained using the elbow method, which gives a visual estimation of the minimum amount of sensors needed for minimum reconstruction error uncertainty. This is followed up by the reconstruction results when applying lcSVD to the data measured by the optimal sensor calculated at the begging of this section compared to the solution obtained when reconstructing uniformly distributed data. Finally, the lcSVD performance metrics are calculated, which consist in the computational cost and the memory consumption. Both OS-lcSVD and lcSVD performance metrics are compared to standard SVD, demonstrating that lcSVD provides better results with a lower computational cost and memory consumption.

\subsection{Optimal sensor placement and data reconstruction using OS-lcSVD \label{Sol_OS-lcSVD}}
The optimal sensor placement for the minimum number of sensors for each dataset has been calculated by utilizing the OS-lcSVD algorithm. In order to test the OS-lcSVD algorithm's reconstruction performance, the number of SVD modes retained to reconstruct the data and validate minimum error is set to $10\%$, $20\%$, $50\%$ and $100\%$ of the minimum number of sensors for each test case. The minimum number of sensors for each dataset is calculated by following the first estimation case presented in Sec.  \ref{sec:elbow}, which, applied to this case, consists of looping through a range of numbers starting at 10, which correspond to the minimum number of sensors the algorithm accepts since the number of retained SVD modes corresponding to $10\%$ of this number is 1, and then calculating the reconstruction $RRMSE$, which must be inferior to a set tolerance $\varepsilon$. The optimal number of sensors is calculated by iterating over a given number of sensors and comparing the reconstruction $RRMSE$ (eq. \eqref{eq:rrmse}), starting off with 10 sensors, in increments of 5 for each step of the iteration. A number of sensors is considered optimal when the reconstruction error decrease stalls for the next set of sensors. The error tolerance $\varepsilon$ is then calculated by rounding the average reconstruction $RRMSE$ obtained after running each case 100 times, since the $RRMSE$ slightly varies for each iteration, to the nearest $0.5$. 

Section \ref{sec:elbow} also presented an alternative method to calculate the optimal number of sensors based on the evolution of the reconstruction error uncertainty. This technique, which is known as the elbow method, is applied when an elbow is formed in the graph with the inflection point corresponding to the optimal number of sensors. From this point onward, uncertainty stops descending at the original rate, suffering a considerable stall. 

\begin{figure}[H]
    \centering
    \includegraphics[width=1\textwidth, angle=0]{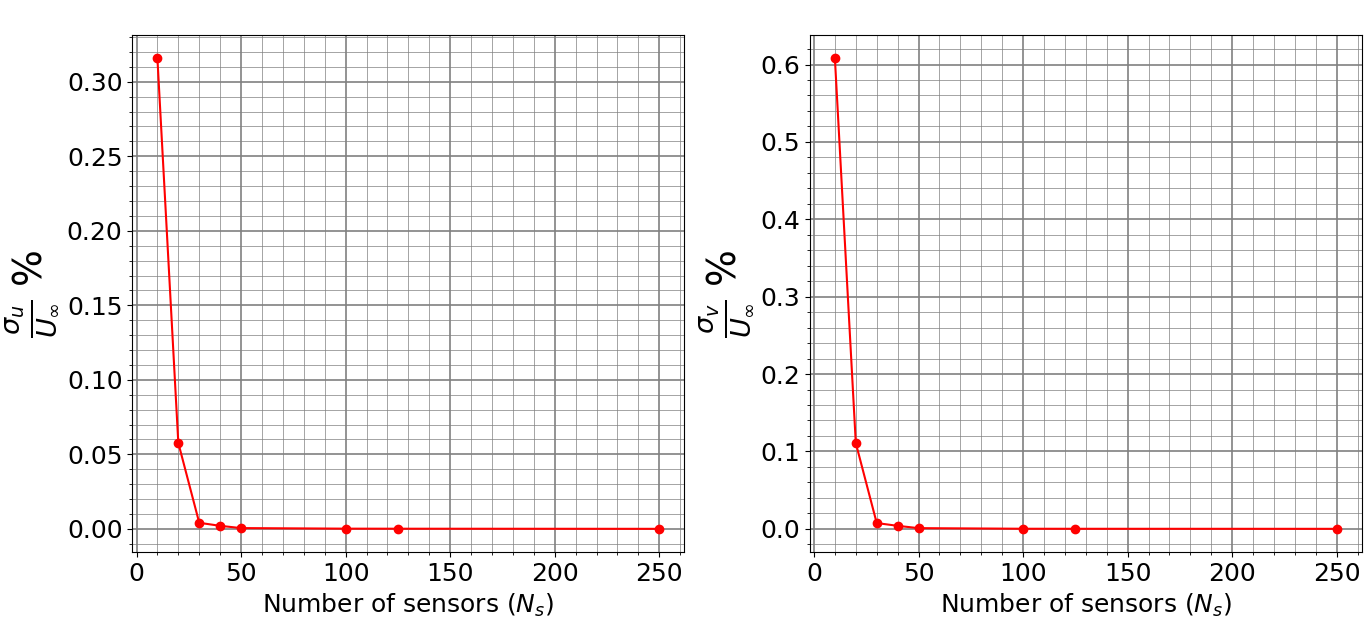}
    \caption{Evolution of the reconstruction error uncertainty ($\sigma$) of the streamwise (left) and normal (right) velocities of the two-dimensional Re $ = 100$ flow cylinder for retained SVD modes equal to 20\% of $N_{s}$.}
    \label{fig:optsenscyl2d}
\end{figure}

This technique works when applied to simple data, such as the two-dimensional Re $= 100$ cylinder. Figure \ref{fig:optsenscyl2d} shows that the optimal number of sensors for the two-dimensional laminar flow cylinder is $N_{s} = 30$, taking into account both uncertainty evolution curves, one for each velocity component.

In this case, adding more sensors does not have a significant impact on the reconstruction error reduction. 

\begin{figure}[H]
    \centering
    \includegraphics[width=1\textwidth, angle=0]{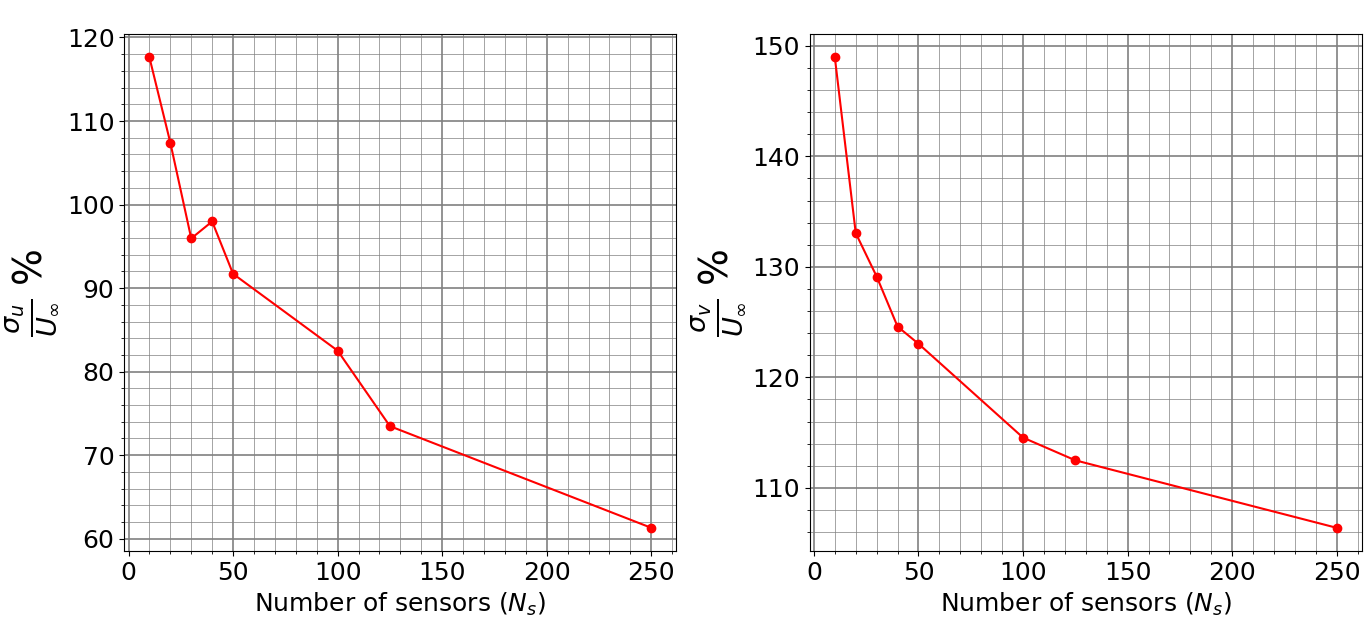}
    \caption{Evolution of the reconstruction error uncertainty ($\sigma$) of the streamwise (left) and normal (right) velocities of the turbulent Re $ = 4000$ flow cylinder for retained SVD modes equal to 20\% of $N_{s}$.}
    \label{fig:optsensvki4000}
\end{figure}

On the other hand, when this technique is applied to more complex data, such as the two-dimensional Re = 4000 turbulent flow cylinder, the elbow does not form where it is expected to do so (at $N_s = 40$). See Fig. \ref{fig:optsensvki4000}.

The uncertainty is much higher than the previous case since the reconstruction error is very high due to this dataset having a large amount of noise (turbulence). In this case, the reconstruction error uncertainty does not stall, but continues to descend. Therefore, the elbow method can not be applied on this type of datasets.

Table \ref{tab:oslcsvdparams} gathers the optimal number of sensors, retained SVD modes, the error tolerance, and the $RRMSE$ for the reconstruction of each test case when $20\%$ of the number of sensors is set as the number of SVD modes to retain.

\begin{table}[H]
    \centering
    \begin{tabular}{|l  l  l  l l|}
    \hline
    \rowcolor{Gray}
    \hline
    \textbf{Dataset} & \textbf{Sensors} & \textbf{SVD modes} & \textbf{$\varepsilon$} & \textbf{$RRMSE$ (\%)} 
    \\ \hline \hline
    2D Cyl100   &   35   & 7 & 1.50 & 1.36063
    \\ 
    3D Cyl280 &   45 & 9       & 4.50 & 4.09697
    \\
    BLayer &   10                 & 2       & 0.50 & $2.466e^{-12}$
    \\
    LES jet     &   10 & 2 & 0.50 & 0.12254
    \\
    2D Cyl4000     &   40 & 8 & 18.00 & 16.9420
    \\
    2D Cyl2600  &   40   & 8 & 16.00 & 15.9305
    \\ 
    
    \hline
\end{tabular}
\caption{Optimal (minimum) number of sensors, number of retained SVD modes for sensor verification via reconstruction, error tolerance $\varepsilon$, and $RRMSE$ for each test case when the number of modes retained is $20\%$ of the number of sensors. \textit{2D Cyl100} represents the Re $ = 100$ 2D cylinder, \textit{3D Cyl280} the Re $ = 280$ 3D cylinder, while \textit{BLayer} represents the Boundary Layer test case, \textit{LES jet} the jet test case, and \textit{2D Cyl4000} and \textit{2D Cyl2600} denote the 2D cylinders at Re = 4000 and Re = 2600, respectively   \label{tab:oslcsvdparams}}
\end{table}

\begin{figure}[H]
    \centering
    \includegraphics[width=1\textwidth, angle=0]{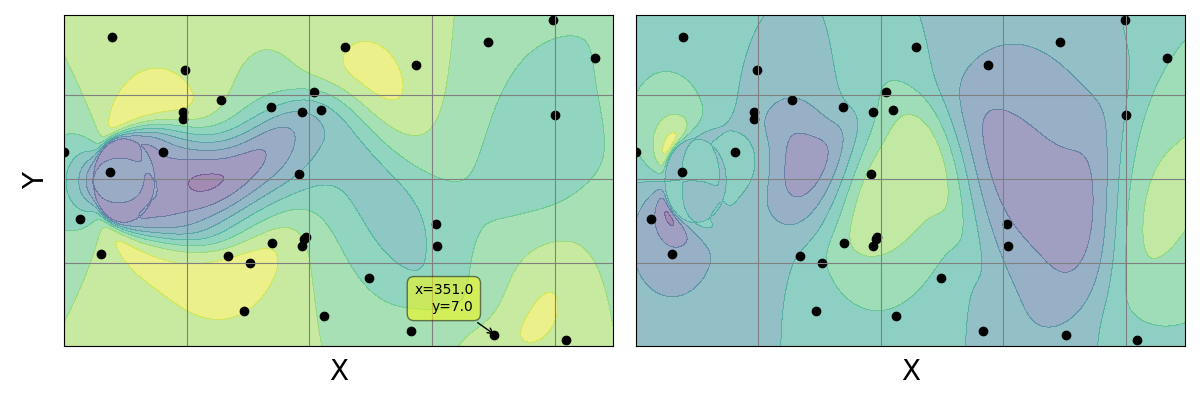}
    \caption{Optimal sensor placement for the two-dimensional Re $ = 100$ cylinder dataset.}
    \label{fig:sensorsCYL2D}
\end{figure}

The two-dimensional Re $= 100$ cylinder requires a higher number of sensors since the dataset, despite lacking data complexity, has a large amount of data points to consider. Notice that the optimal number of sensors via iteration is $N_s = 35$, while the estimated number using the elbow method is $N_s = 30$, proving that this second method gives a close estimation. It is important to note that this dataset has two velocity components, so the sensors are positioned to capture the most important data taking into account both of these. In other words, the sensor positions are the same for each velocity components when measured in the same spatial plane. Figure \ref{fig:sensorsCYL2D} displays the optimal sensor positions for the two-dimensional laminar flow cylinder. Notice that in the figure, the $mplcursor$ library has been used to create interactive plots, allowing users to hover over each sensor and display the sensor's coordinates.

\begin{figure}[H]
    \centering
    \includegraphics[width=1\textwidth, angle=0]{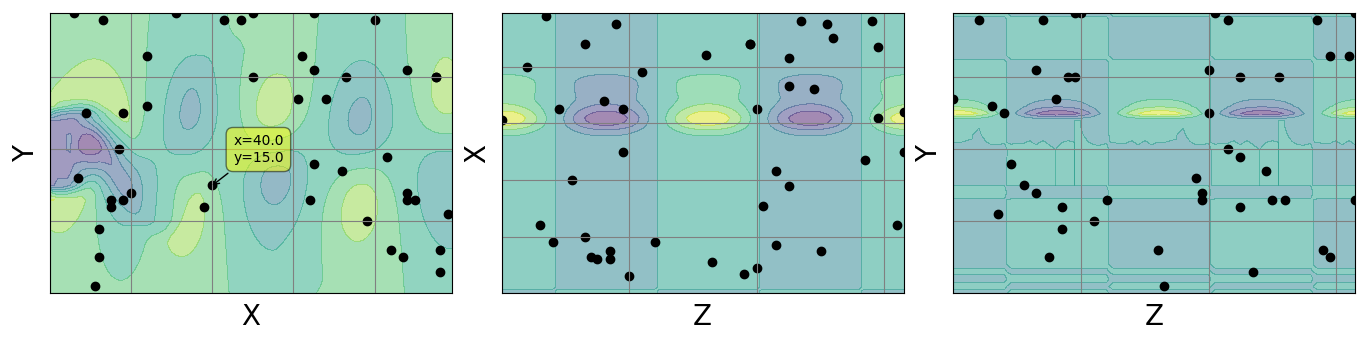}
    \caption{Optimal sensor placement in each one of the planes of the three-dimensional Re $ = 280$ cylinder dataset.}
    \label{fig:sensorsCYL3D}
\end{figure}

The three-dimensional Re $= 280$ cylinder requires the highest number of sensors since it has three dimensions and, therefore, three velocity components, meaning more data needs to be collected. Figure \ref{fig:sensorsCYL3D} displays the sensor positions for each one of the spatial planes.

The previous datasets can be reconstructed with a higher amount of retained SVD modes since both are laminar, so there is a lack of noise in the data.

A low reconstruction error is obtained for both the turbulent boundary layer and LES jet test cases when using the minimum number of sensors possible and reconstructing the datasets with 2 retained singular values. This is due to the number of retained singular values being proportional to the number of sensors. Since the data is turbulent, a low number of retained singular values transfers to the maximum noise possible being filtered out (representing small uncorrelated flow scales) during reconstruction. At the same time, the number of sensors is low since both datasets have less data points in comparison to the rest of test cases, and the most relevant information in concentrated in the lower half of the spatial grid.

The last two test cases, which are the two-dimensional experimental datasets representing the turbulent Re $= 4000$ and Re $= 2600$ cylinders, require the same number of sensors for optimal reconstruction. This is mainly due to both datasets containing the same number of data points and velocity components. The difference between reconstruction $RRMSE$'s lays on the higher turbulence level (acting in the similar way as noise) contained in the Re $= 4000$ case. These cases demonstrate the importance of the balance between the number of sensors and retained SVD modes. In these cases, a greater number of sensors is required given the dimensions of the spatial mesh but, at the same time, a proportional increase of the retained SVD modes translates to a larger amount of noise being retained during reconstruction. 

\begin{figure}[H]
    \centering
    \includegraphics[width=1\textwidth, angle=0]{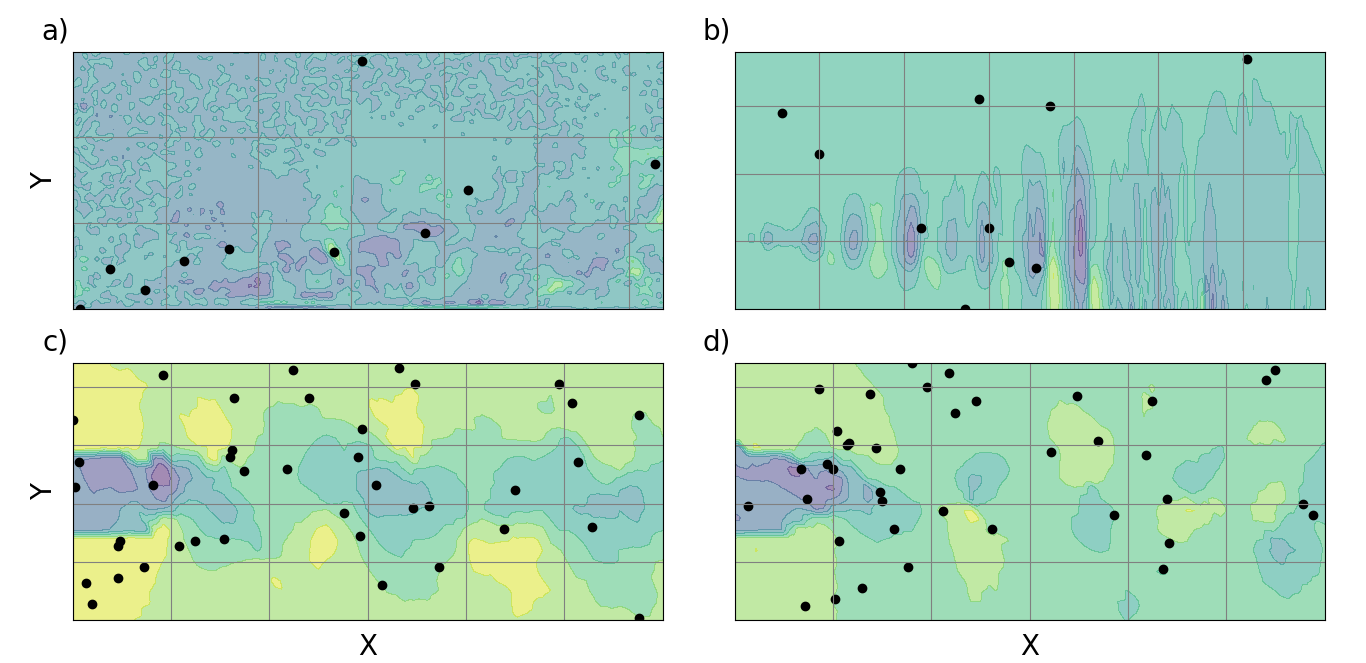}
    \caption{Optimal sensor placement for: a) The turbulent boundary layer test case, b) the turbulent jet LES, c) the two-dimensional Re = 4000 cylinder and d) the two-dimensional Re = 2600 cylinder. The sensors are represented on the streamwise velocity.}
    \label{fig:OSothers}
\end{figure}

Figure \ref{fig:OSothers} presents the optimal sensor positions for the remaining test cases. 

\begin{figure}[H]
    \centering
    \includegraphics[width=1\textwidth, angle=0]{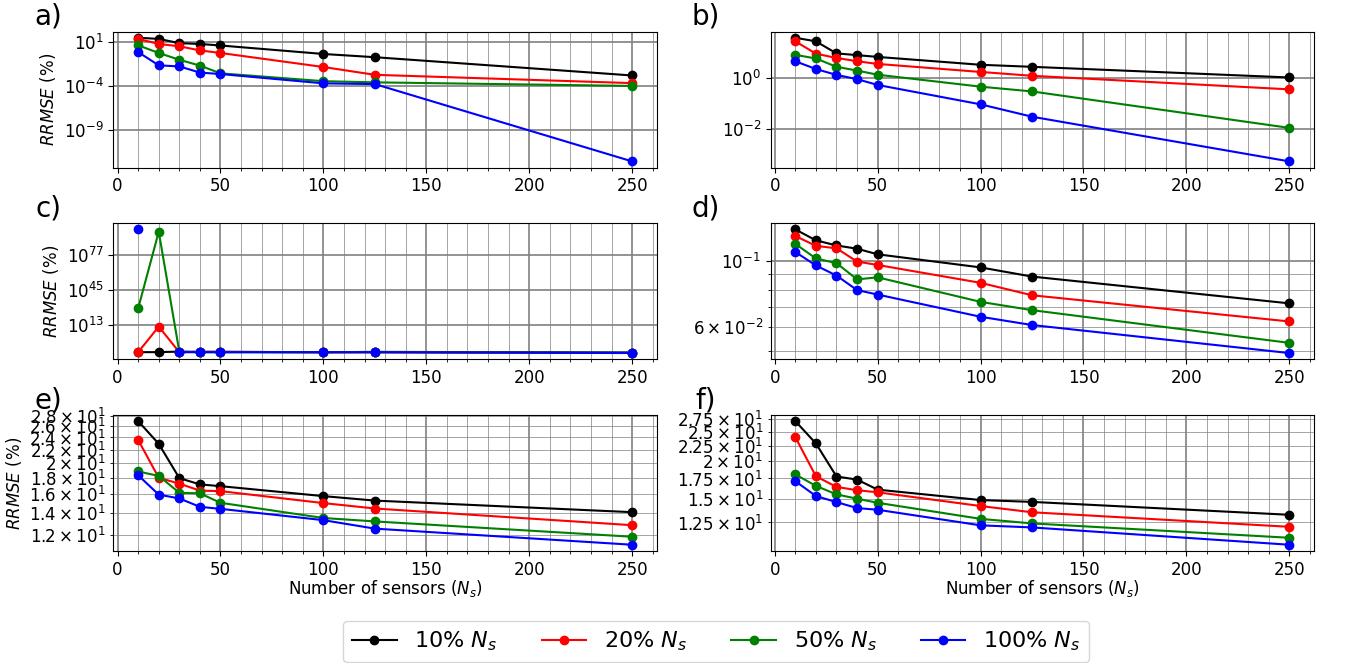}
    \caption{Reconstruction $RRMSE$ evolution of all retained SVD modes cases for: a) The two-dimensional Re $ = 100$ cylinder, b) Three-dimensional Re $ = 280$ cylinder, c) turbulent boundary layer test case, d) the turbulent jet LES, e) the turbulent Re = 4000 cylinder and f) the turbulent Re = 2600 cylinder. The curves represent retained SVD modes equal to 10\%, 20\%, 50\%, and 100\% of the number of sensors $N_s$.
    \label{fig:osresultsall}}
\end{figure}

Despite presenting the results for the number of retained SVD modes equal to 20\% of $N_s$, all optimal number of sensors estimations for each one of the retained SVD modes cases, for each dataset, are presented in Fig. \ref{fig:osresultsall}.

The reconstruction $RRMSE$ curves for each number of retained SVD modes seem to stack in the same order for all test cases. The curve that represents the error values when retaining a number of SVD modes equal to 10\% of the number of sensors represents the highest error value for each number of sensors, while the opposite effect occurs with the curve that represents retained SVD modes equal to 100\% of the number of sensors. Also, the error decreases when the number of sensors increases. This decrease in the $RRMSE$ value in more noticeable in datasets that contain laminar data rather than turbulent data, which is related to the data complexity.

 Table \ref{tab:sensorlcsvdsol} displays the original dataset shape presented in Sec. \ref{sec:database}, a reorganized version of each dataset following Sec. \ref{sec:dataOrganization}, where the dimensions are reshaped to $J \times K$, with $J$ containing the compressed spatial data, and the dimensional information of the dataset generated by the sensors with dimensions $N_{s} \times K$, where $N_{s}$ is the number of sensors calculated in Tab. \ref{tab:oslcsvdparams}. The last column of this table contains the spatial data compression rate $C_{R}$ comparing the original number of spatial points and the number of sensors selected as 
\begin{equation}
    C_{R} = \frac{J}{N_{s}}.
    \label{eq:compRate}
\end{equation}

\begin{table}[H]
    \centering
    \begin{tabular}{|l  l  l  l l|}
    \hline
    \rowcolor{Gray}
    \hline
    \textbf{Dataset} & \textbf{Shape} & \textbf{$J \times K$} & \textbf{$N_{s} \times K$} & \textbf{$C_{R}$} 
    \\ \hline \hline
    2D Cyl100  & $2 \times 449 \times 199 \times 151$ & $1787025 \times 151$ & $35 \times 151$ & $51058$
    \\
    3D Cyl280 & $3 \times 100 \times 40 \times 64 \times 299$ & $768000 \times 299$ & $45 \times 299$ & $17066$
    \\
    BLayer &   $319 \times 150 \times 6170$                 & $47850 \times 6170$       & $10 \times 6170$ & $4785$
    \\
    LES jet     &   $175 \times 39 \times 5000$ & $6825 \times 5000$ & $10 \times 5000$ & $683$
    \\
    2D Cyl4000 & $2 \times 301 \times 111 \times 4000$ & $33411 \times 4000$ & $40 \times 4000$ & $835$
    \\
    2D Cyl2600 & $2 \times 301 \times 111 \times 5200$ & $33411 \times 5200$ & $40 \times 5200$ & $835$
    \\
    
    \hline
\end{tabular}
\caption{From left to right columns: Dataset shape, reorganized data shape, shape of the data collected by the optimal sensors, and the data compression rate between the original data and the sensor dataset. The dataset shape is $N_x \times N_y \times K$ for two-dimensional data with a singular velocity component, $N_{comp} \times N_x \times N_y \times K$ for two-dimensional data with more than one component, and $N_{comp} \times N_x \times N_y \times N_z \times K$, for three-dimensional data with multiple velocity components. $C_{R}$ is the spatial data compression rate calculated using eq. \eqref{eq:compRate}. \label{tab:sensorlcsvdsol}}
\end{table}

Notice how, for the two-dimensional Re $= 100$ cylinder, data is compressed from 1787025 spatial data points to 35, meaning the spatial data has been compressed by 51058. This means that the data collected by 35 optimal sensors is sufficient to reconstruct the remaining 1786990 data points. High compression rates are also achieved with turbulent data, which is highly complex, such as the turbulent Re $= 4000$ and Re $= 2600$ cylinder which both share the same compression rate of 835 given that their spatial dimensions are the same, as well as the optimal number of sensors.

\begin{figure}[H]
    \centering
    \includegraphics[width=1\textwidth, angle=0]{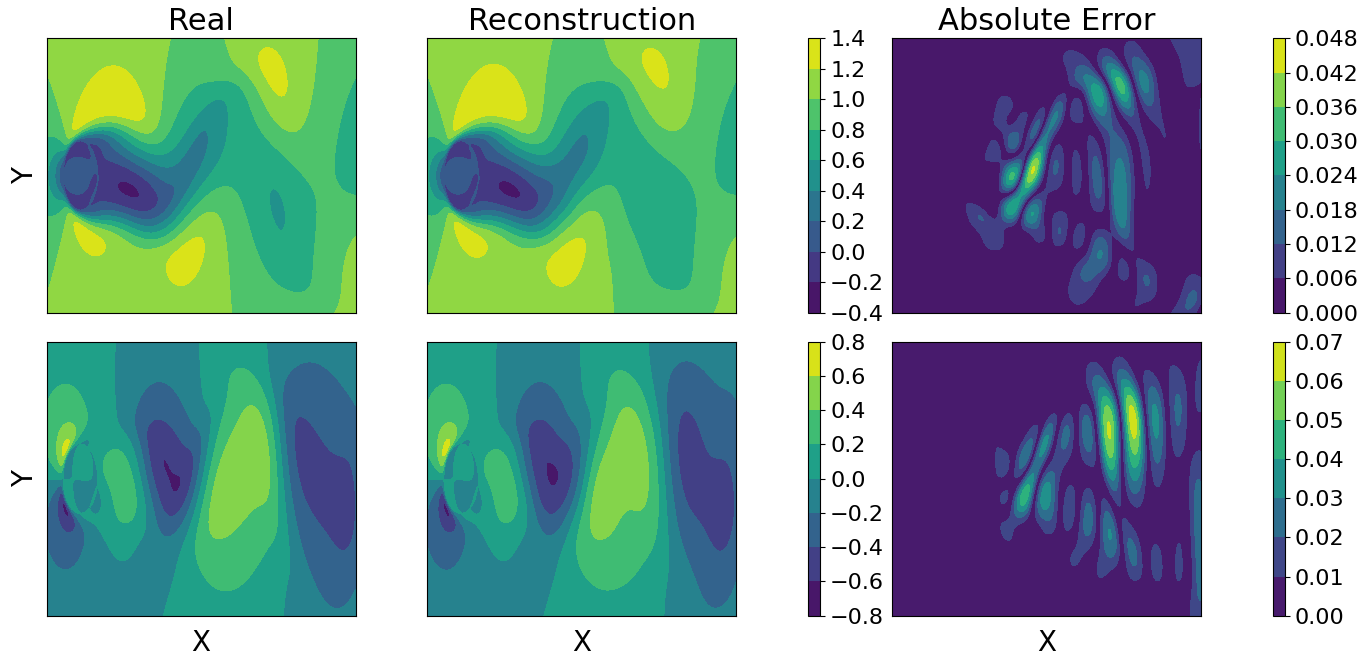}
    \caption{OS-lcSVD reconstruction results for the two-dimensional Re $ = 100$ cylinder. From left to right and top to bottom: the real data, the reconstructed data, and the absolute error for the highest error snapshot of the streamwise velocity, and below are the same results for the normal velocity.}
    \label{fig:CYL2Dreconst}
\end{figure}

The dataset reconstruction results for the two-dimensional Re $=100$ cylinder can be seen in Fig. \ref{fig:CYL2Dreconst}, where a comparison between the original dataset, the solution dataset after OS-lcSVD completes the reconstruction using the data collected by the optimal sensors $N_s$, and the reconstruction error contour map for both velocity components for the highest reconstruction error snapshot are presented.

The reconstruction error contour maps from Fig.\ref{fig:CYL2Dreconst} show that the reconstruction error scale is $10^{-2}$, meaning the reconstruction using a optimally placed minimum number of sensors is very precise.

\begin{figure}[H]
    \centering
    \includegraphics[width=1\textwidth, angle=0]{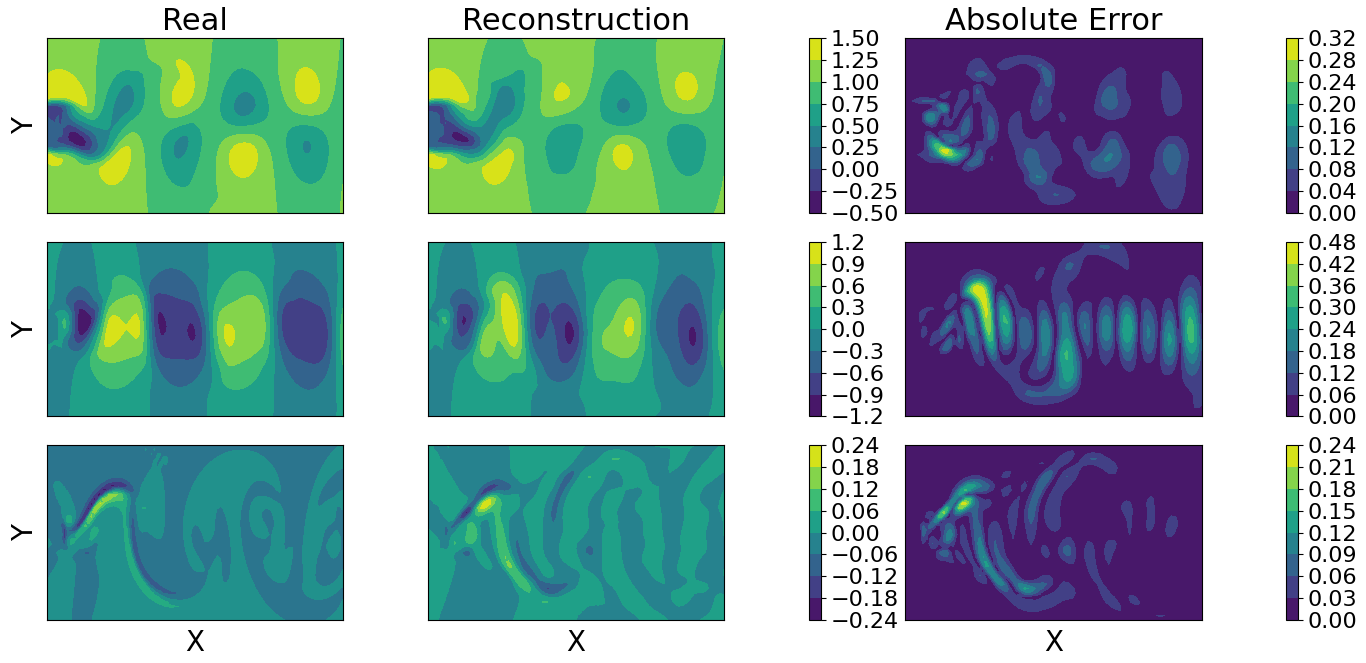}
    \caption{OS-lcSVD reconstruction results for the three-dimensional Re $ = 280$ cylinder. From left to right and top to bottom: the real data, the reconstructed data, and the absolute error for the highest error snapshot of the streamwise velocity, and in the middle row are the same results for the normal velocity, while spanwise velocity results are presented in the bottom row.}
    \label{fig:CYL3Dreconst}
\end{figure}

The dataset reconstruction results for the three-dimensional Re $ = 280$ are presented in Fig. \ref{fig:CYL3Dreconst} for the maximum error plane and snapshot of each velocity component.

The absolute error values shown in the error contour maps of Fig. \ref{fig:CYL3Dreconst} are mostly 0. For the spanwise velocity, a small amount of peak error values match the velocity values. This is caused due to the velocity values for this component being the same scale as the absolute error ($10^{-1}$), and this error scale depends on the selected parameters (number of sensors and retained singular values). Therefore, this can be solved by increasing the number of sensors or retained SVD modes, as seen in Fig. \hyperref[fig:OSothers]{\ref*{fig:OSothers}(b)}. 

\begin{figure}[H]
    \centering
    \includegraphics[width=1\textwidth, angle=0]{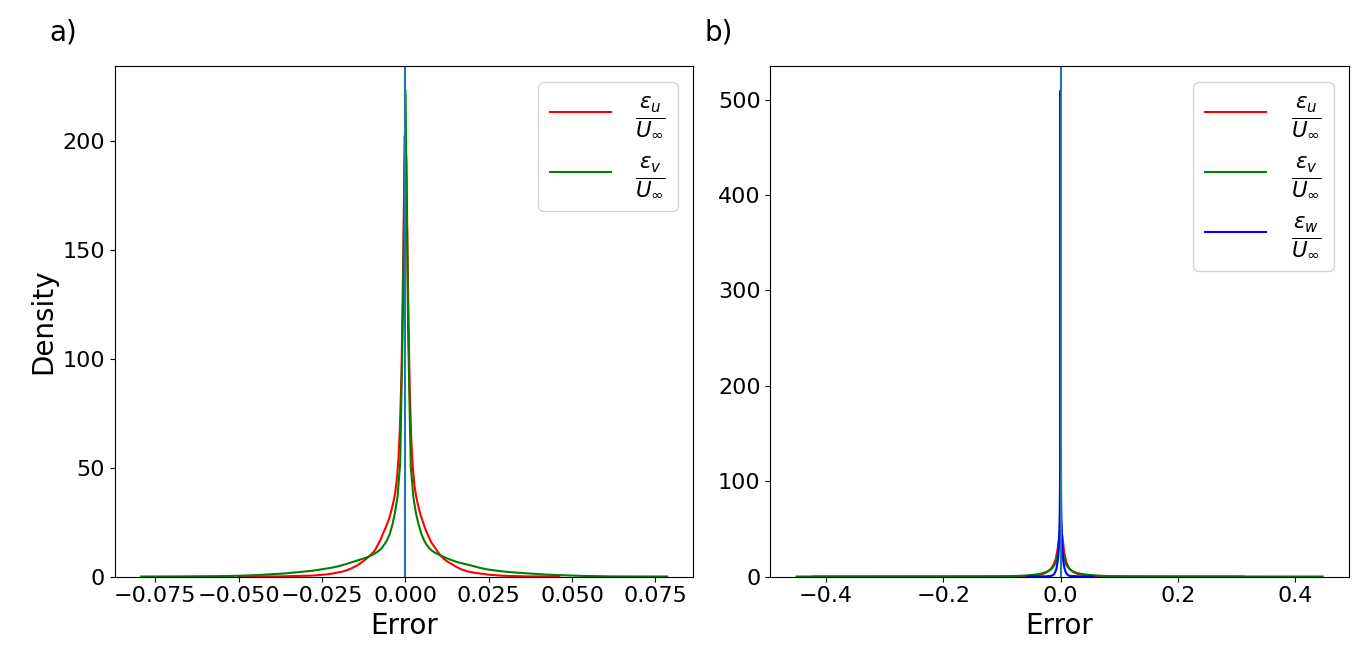}
    \caption{OS-lcSVD reconstruction error probability density curves for: a) the streamwise and normal velocity components of the two-dimensional Re $= 100$ cylinder and b) the streamwise, normal and spanwise velocity components of the three-dimensional Re $= 280$ cylinder.}
    \label{fig:CYL2D3Dpdf}
\end{figure}

Figure \ref{fig:CYL2D3Dpdf} illustrates the reconstruction error probability density curve of each one of the velocity components of both the two- and three-dimensional laminar flow cylinder datasets.

The error scale of the two-dimensional Re $ = 100$ cylinder is in the order of $10^{-2}$, while for the three-dimensional Re $ = 280$ cylinder, it is in the order of $10^{-1}$. The horizontal axis contains the error values, while the vertical axis represents the density or frequency of said values, forming what is known as the density curve. The density curves for both test cases and all velocity components are narrow and centered in 0, which means that the majority of reconstruction error values are 0, while the number of outlines is minimum. The density curves also give information about the probability of a reconstruction error being of a determined value since they inform about the error data distribution. The previously described information can be visually contrasted with the reconstruction error contour maps previously presented in Fig. \ref{fig:CYL2Dreconst} and Fig. \ref{fig:CYL3Dreconst}.

\begin{figure}[H]
    \centering
    \includegraphics[width=1\textwidth, angle=0]{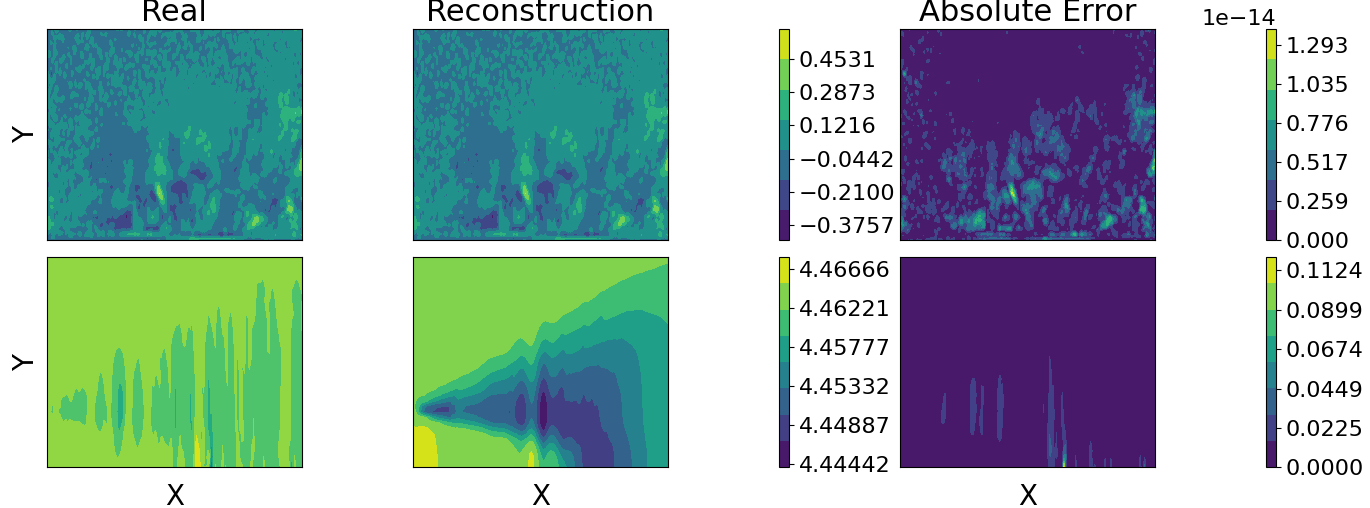}
    \caption{OS-lcSVD reconstruction results. From left to right and top to bottom: the real data, the reconstructed data, and the absolute error for the highest error snapshot of the turbulent boundary layer streamwise velocity, and these same results for the turbulent jet LES in the bottom row.}
    \label{fig:groupedOSLCSVD_1}
\end{figure}

\begin{figure}[H]
    \centering
    \includegraphics[width=1\textwidth, angle=0]{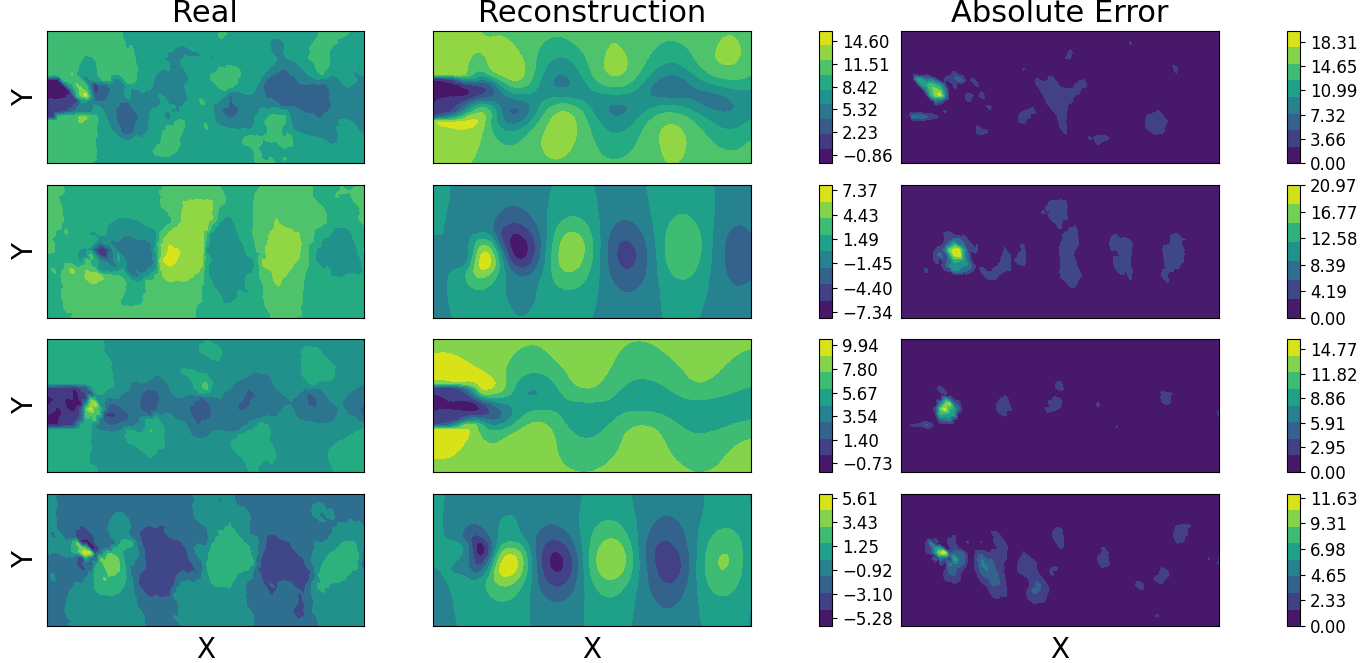}
    \caption{OS-lcSVD reconstruction results. From left to right and top to bottom: the real data, the reconstructed data, and the absolute error for the highest error snapshot of the streamwise and spanwise velocities of the turbulent Re $= 4000$ are presented in the first two rows, and the same results for the Re $= 2600$ cylinder's streamwise and normal velocities are presented in the last two rows.}
    \label{fig:groupedOSLCSVD_2}
\end{figure}

Figure \ref{fig:groupedOSLCSVD_1} illustrates the reconstruction results for the turbulent boundary layer and jet, while Fig. 
\ref{fig:groupedOSLCSVD_2} presents the results for both two-dimensional turbulent flow cylinder datasets. Both figures present the results for the snapshot that contains the highest absolute error for each velocity component.

The turbulent boundary layer absolute error (Fig. \ref{fig:groupedOSLCSVD_1}) is in the order of $10^{-14}$, which is expected based on the reconstruction $RRMSE$ result presented in Tab. \ref{tab:oslcsvdparams}. The bottom row of this same figure shows the absolute error of the turbulent jet dataset. Note that both datasets are turbulent. This proves the lc-SVD algorithm's high capability of filtering noise to produce precise dataset reconstructions. 

The small number of peak absolute error values appreciated in Fig. \ref{fig:groupedOSLCSVD_2} for both components of the turbulent Re $= 4000$ and Re $=2600$. These  error values have several causes, being the complexity of the data, the number of sensors, and the number of retained SVD modes, with the last two being leveraged in order to obtain the lowest reconstruction error possible. Since these few peak values are sparsely distributed along the spatial mesh, and taking into account that the absolute error results presented in Fig. \ref{fig:groupedOSLCSVD_2} are the highest among the full dataset, it can be determined that the overall reconstruction is highly accurate.

\begin{figure}[H]
    \centering
    \includegraphics[width=1\textwidth, angle=0]{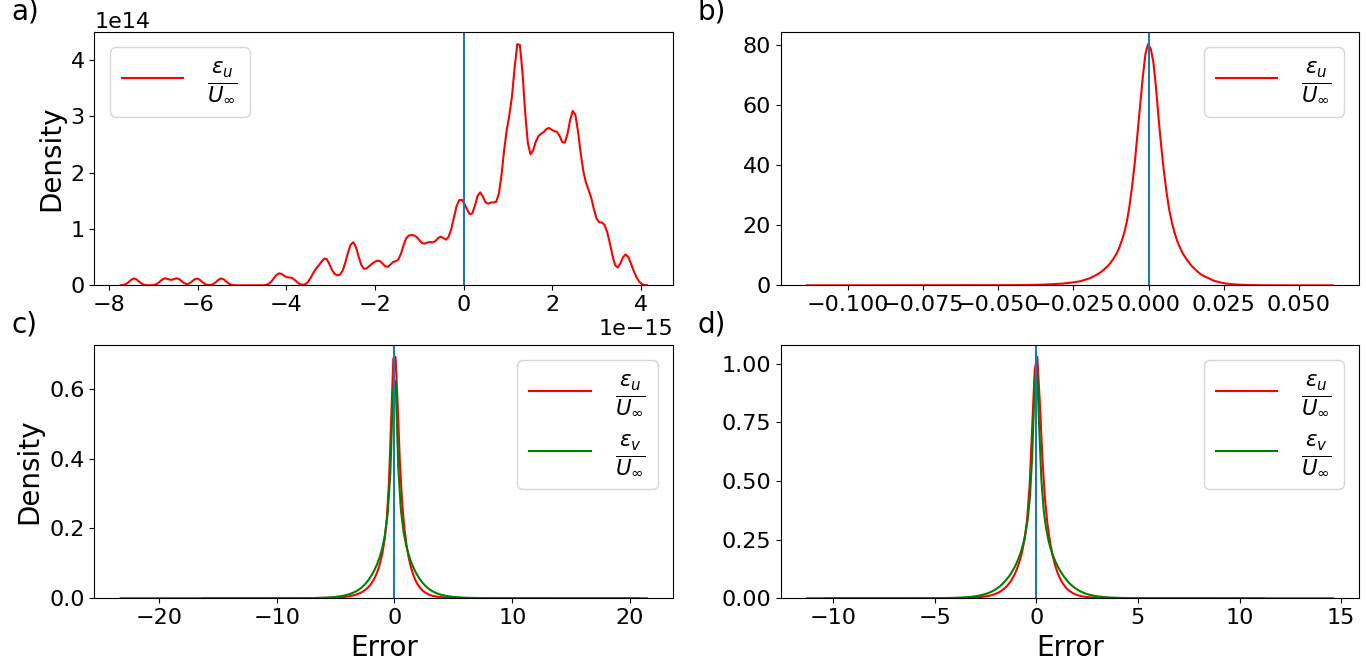}
    \caption{OS-lcSVD reconstruction error probability density curves for: a) the streamwise velocity of the turbulent boundary layer, b) the streamwise velocity of the turbulent jet LES, c) the streamwise and normal velocity of the turbulent Re $= 4000$ cylinder, and d) the the streamwise and normal velocity of the turbulent Re $= 2600$ cylinder.}
    \label{fig:groupedPDFOSLCSVD}
\end{figure}

The reconstruction error probability density curves for all velocity components of the remaining test cases are shown in Fig. \ref{fig:groupedPDFOSLCSVD}. 

Notice that the density curves for the turbulent boundary layer (Fig. \hyperref[fig:groupedPDFOSLCSVD]{\ref*{fig:groupedPDFOSLCSVD}(a)}) and the turbulent jet LES (Fig. \hyperref[fig:groupedPDFOSLCSVD]{\ref*{fig:groupedPDFOSLCSVD}(b)}) do not follow a normal distribution, but instead a skewed left distribution, meaning that there is an imbalance in the magnitudes of the velocity error values to the left and right of zero (where they should be centered). In addition to this, the turbulent boundary layer data distribution is not centered in zero, but at $10^{-15}$, with this being due to the precision in the error scale in comparison to the other cases. 
The turbulent Re $= 4000$ cylinder (Fig. \hyperref[fig:groupedPDFOSLCSVD]{\ref*{fig:groupedPDFOSLCSVD}(c)}) error data follows a normal distribution, similar to those of the laminar cylinders, while the turbulent Re $= 2600$ cylinder (Fig. \hyperref[fig:groupedPDFOSLCSVD]{\ref*{fig:groupedPDFOSLCSVD}(d)}) follows Poisson distribution since the right tail stretches further than the left tail. Notice that the density scale for the reconstruction errors of these two test cases is much lower than the rest, meaning that there is a broader variety of error values since the reconstruction error value scale is larger.

\subsection{lcSVD applied to equidistant spatial data points \label{Sol_lcSVD}}
Using an amount of equidistantly distributed points in $x$ and $y$ (and $z$ for the three-dimensional cylinder test case), reshaped to $\overline J$, equivalent to the minimum number of sensors $N_{s}$ calculated in Sec. \ref{Sol_OS-lcSVD}, the datasets are then reconstructed and the error is estimated the same way the previous section. The dataset are compressed down to the same shapes as presented in Tab. \ref{tab:sensorlcsvdsol}, and are shown again in Tab. \ref{tab:equidistlcsvdsol}, together with the reconstruction $RRMSE$ for each dataset after applying lcSVD. In order to compare the results with those obtained in the previous results section, the same amount of singular values are retained, which is 20\% of $\overline J$, with $\overline J = N_s$.

\begin{table}[H]
    \setlength{\extrarowheight}{.5ex}
    \centering
    \begin{tabular}{|l l l l|}
    \hline
    \rowcolor{Gray}
    \hline
    \textbf{Dataset} & \textbf{$J \times K$} & \textbf{$\overline J \times K$} & \textbf{$RRMSE$ (\%)} 
    \\ \hline \hline
    2D Cyl100 & $1787025 \times 151$  & $35 \times 151$ & $1.47456$
    \\
    3D Cyl280 & $768000 \times 299$ & $45 \times 299$ & $3.74412$
    \\
    BLayer & $47850 \times 6170$ & $10 \times 6170$ & $6.425e^{-13}$
    \\
    LES jet   & $6825 \times 5000$ & $10 \times 5000$ & $0.12375$
    \\
    2D Cyl4000 & $33411 \times 4000$ & $40 \times 4000$ & $15.8331$
    \\
     2D Cyl2600 & $33411 \times 5200$ & $40 \times 5200$ & $15.0756$
    \\
    \hline
\end{tabular}
\caption{Original dataset shape, reduced shape after selecting equidistant data points, and reconstruction $RRMSE$ using lcSVD for each dataset when selecting equidistant data points. The spatial components $N_{comp} \times N_x \times N_y \;(\times \; N_z)$ are reshaped into $J$ for all $K$ snapshots, and then the dataset dimensions are reduced selecting $\overline J = N_s$ equidistant data points. \label{tab:equidistlcsvdsol}}
\end{table}

\begin{figure}[H]
    \centering
    \includegraphics[width=1\textwidth, angle=0]{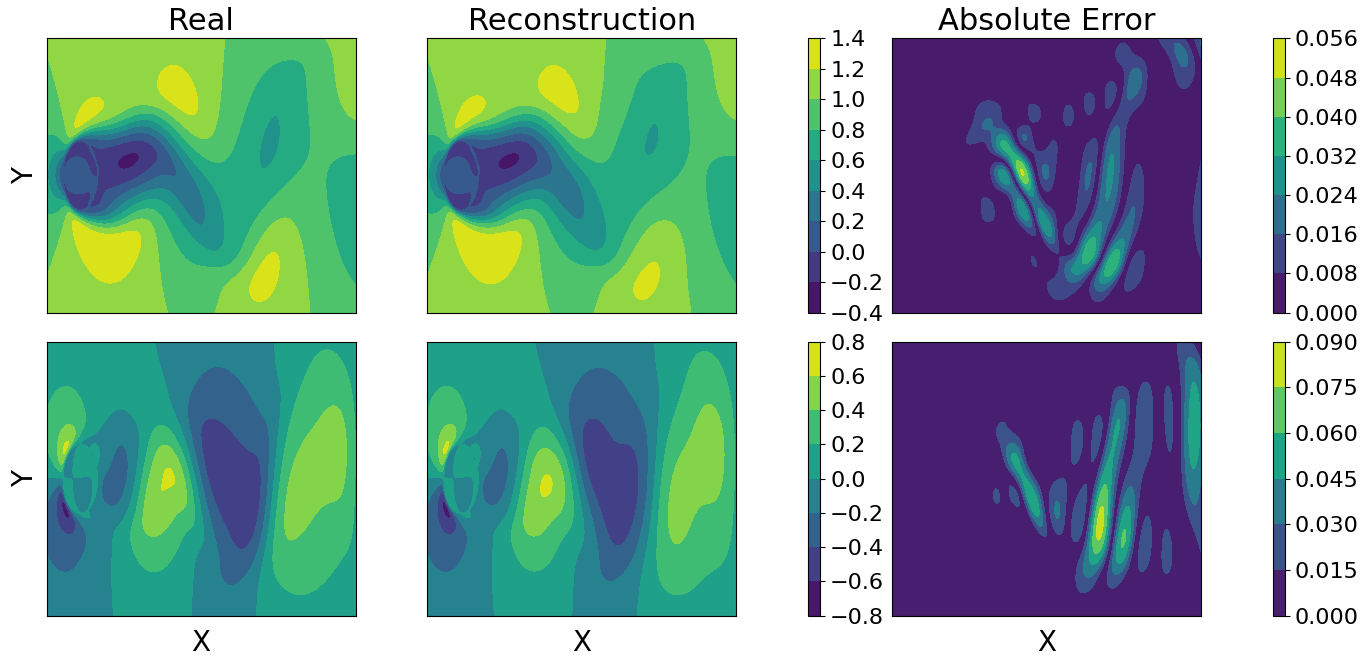}
    \caption{lcSVD reconstruction results for the two-dimensional Re $ = 100$ cylinder. From left to right and top to bottom: the real data, the reconstructed data, and the absolute error for the highest error snapshot of the streamwise velocity, and below are the same results for the normal velocity.}
    \label{fig:eq_CLY2Drec}
\end{figure}

Figure \ref{fig:eq_CLY2Drec} presents the original data, the reconstructed data after applying lcSVD to the reduced equidistant spatial data, and the reconstruction absolute error contour map for the maximum error snapshot of each of the two velocity components of the two-dimensional Re $ = 100$ flow cylinder. 

The absolute error values of both velocity components are similar to those obtained for this same dataset when using optimal sensors, which can be seen in Fig. \ref{fig:CYL2Dreconst}. This demonstrates the lcSVD algorithm's data reconstruction capabilities when using equidistant data. 

\begin{figure}[H]
    \centering
    \includegraphics[width=1\textwidth, angle=0]{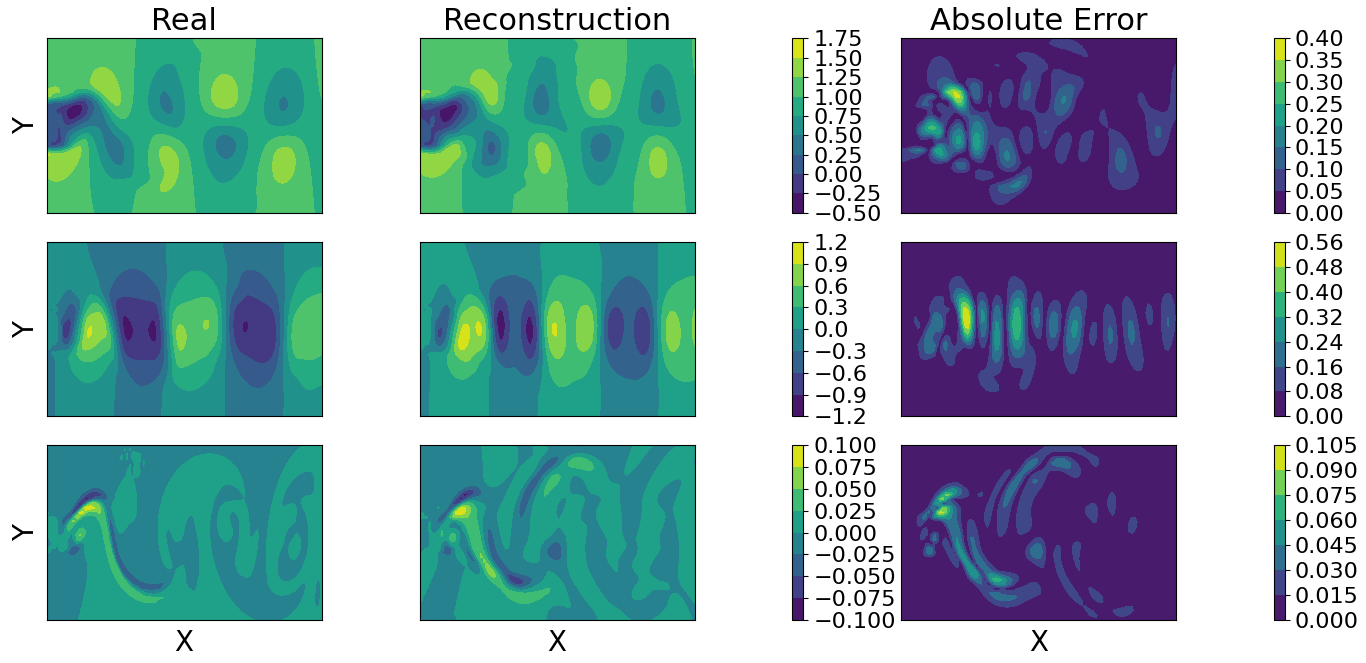}
    \caption{lcSVD reconstruction results for the three-dimensional Re $ = 280$ cylinder. From left to right and top to bottom: the real data, the reconstructed data, and the absolute error for the highest error snapshot of the streamwise velocity, and in the middle row are the same results for the normal velocity, while spanwise velocity results are presented in the bottom row.}
    \label{fig:eq_CLY3Drec}
\end{figure}

The reconstruction comparison for the three-dimensional Re $= 280$ cylinder is presented in Fig. \ref{fig:eq_CLY3Drec}, comparing the original data, reconstruction using lcSVD on equidistant spatial data, and the absolute error contour map for the highest error snapshot and plane of each velocity component.

For the three-dimensional dataset, the absolute error scale for the spanwise velocity once again matches the velocity field scale, with the causes being identical to those described for the reconstruction results of this same dataset using OS-lcSVD in Sec. \ref{Sol_OS-lcSVD}.

\begin{figure}[H]
    \centering
    \includegraphics[width=1\textwidth, angle=0]{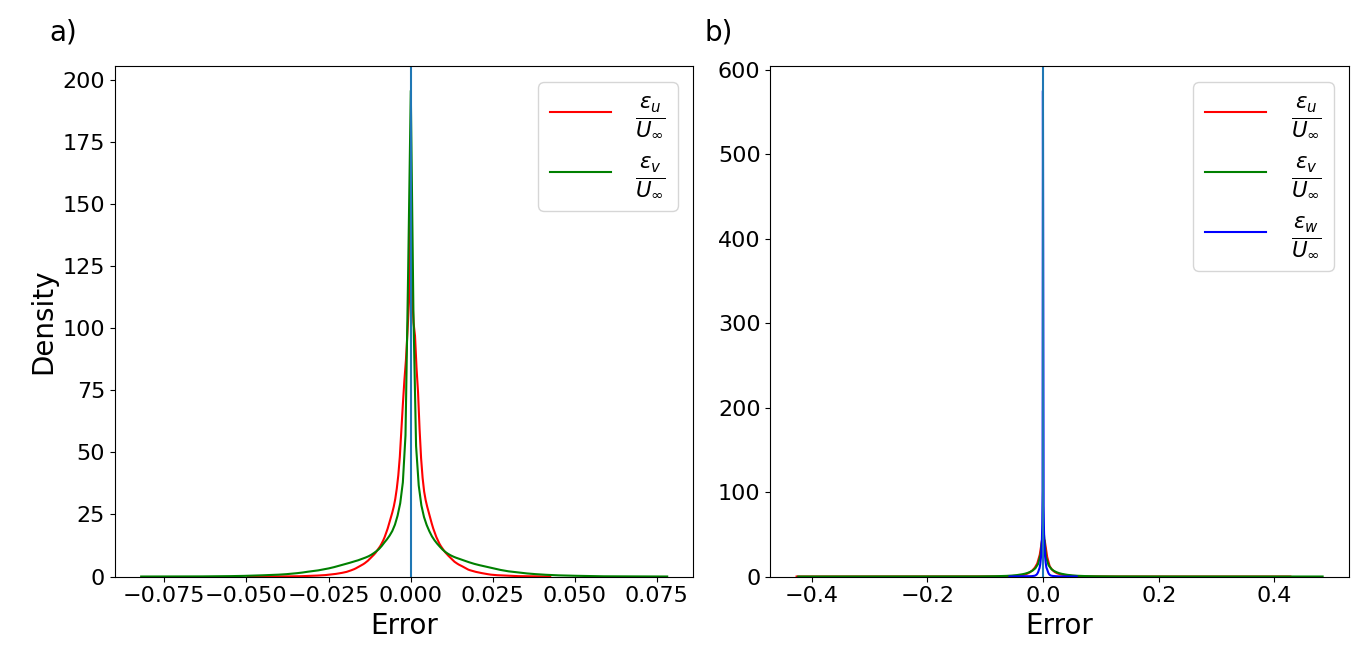}
    \caption{lcSVD reconstruction error probability density curves for equidistant data: a) the streamwise and normal velocity components of the two-dimensional Re $= 100$ cylinder and b) the streamwise, normal and spanwise velocity components of the three-dimensional Re $= 280$ cylinder.}
    \label{fig:eqCYL2D3Dpdf}
\end{figure}

The reconstruction error uncertainty for the previous datasets is illustrated in Fig. \ref{fig:eqCYL2D3Dpdf}, where the probability density curves of all velocity components of these two datasets can be seen.

For the two-dimensional Re $= 100$ cylinder (Fig. \hyperref[fig:eqCYL2D3Dpdf]{\ref*{fig:eqCYL2D3Dpdf}(a)}), the error values are similar to those registered in Fig. \hyperref[fig:CYL2D3Dpdf]{\ref*{fig:CYL2D3Dpdf}(a)}, but the density for error value 0 is lower. The opposite effect occurs with the three-dimensional Re $= 280$ cylinder, Fig. \hyperref[fig:eqCYL2D3Dpdf]{\ref*{fig:eqCYL2D3Dpdf}(b)}. This is directly related to the reconstruction $RRMSE$ for each dataset, which are displayed in Tab. \ref{tab:oslcsvdparams} for optimal sensors, and Tab. \ref{tab:equidistlcsvdsol} for equidistant spatial data. 

\begin{figure}[H]
    \centering
    \includegraphics[width=1\textwidth, angle=0]{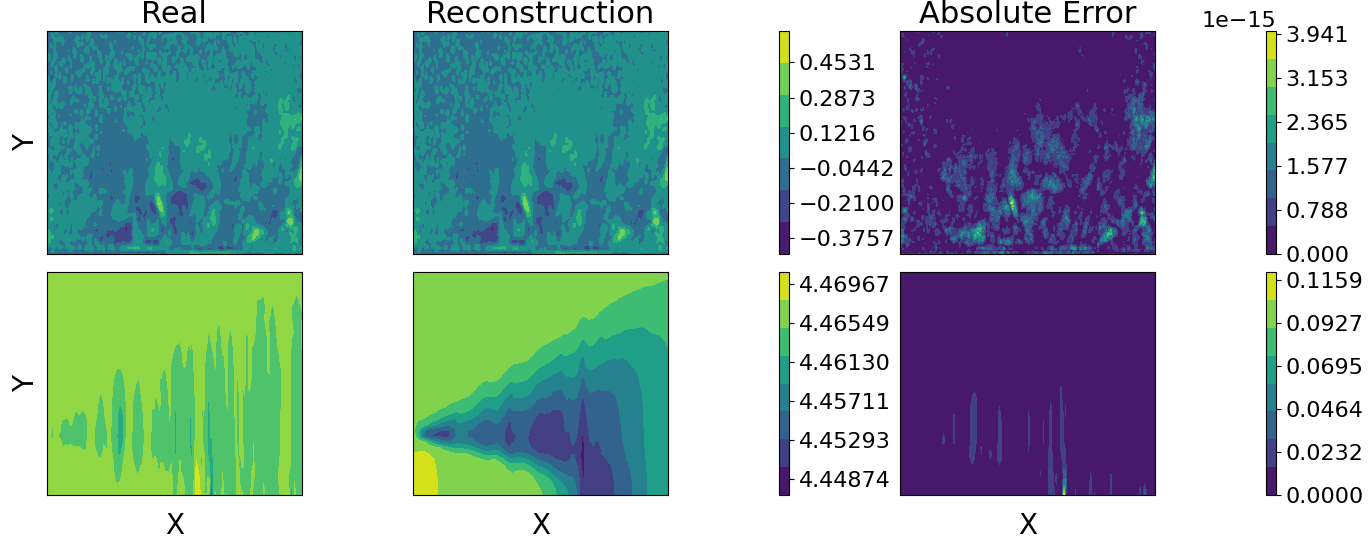}
    \caption{lcSVD reconstruction results for equidistant data. From left to right and top to bottom: the real data, the reconstructed data, and the absolute error for the highest error snapshot of the turbulent boundary layer streamwise velocity, and these same results for the turbulent jet LES in the bottom row.}
    \label{fig:groupedLCSVD_1}
\end{figure}

\begin{figure}[H]
    \centering
    \includegraphics[width=1\textwidth, angle=0]{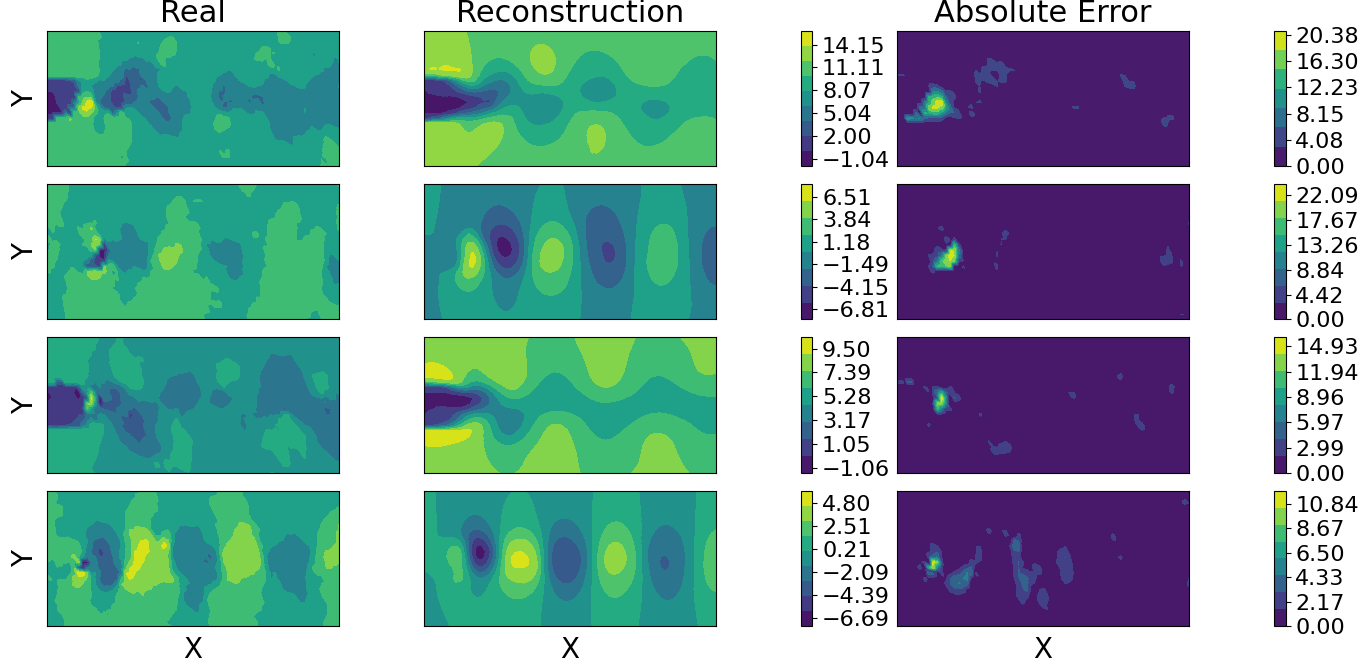}
    \caption{lcSVD reconstruction results for equidistant data. From left to right and top to bottom: the real data, the reconstructed data, and the absolute error for the highest error snapshot of the streamwise and spanwise velocities of the turbulent Re $= 4000$ are presented in the first two rows, and the same results for the Re $= 2600$ cylinder's streamwise and normal velocities are presented in the last two rows.}
    \label{fig:groupedLCSVD_2}
\end{figure}

The data reconstruction results using lcSVD on equidistantly distributed spatial data of the turbulent boundary layer and jet LES can be seen in Fig. \ref{fig:groupedLCSVD_1}, while the results for both turbulent Re $= 4000$ and Re $= 2600$ are presented in Fig. \ref{fig:groupedLCSVD_2}.

Once again, the reconstruction $RRMSE$ errors presented in Tab. \ref{tab:oslcsvdparams} for optimal sensors are similar to those gathered in Tab. \ref{tab:equidistlcsvdsol}, meaning that lcSVD works well when reconstructing data using the data collected by optimally placed sensors, or with equidistantly distributed spatial data. Despite this, it is recommended to use optimal sensors to guarantee maximum reconstruction precision, since there is a possibility that using equidistant data will not always give as precise reconstruction results.

\begin{figure}[H]
    \centering
    \includegraphics[width=1\textwidth, angle=0]{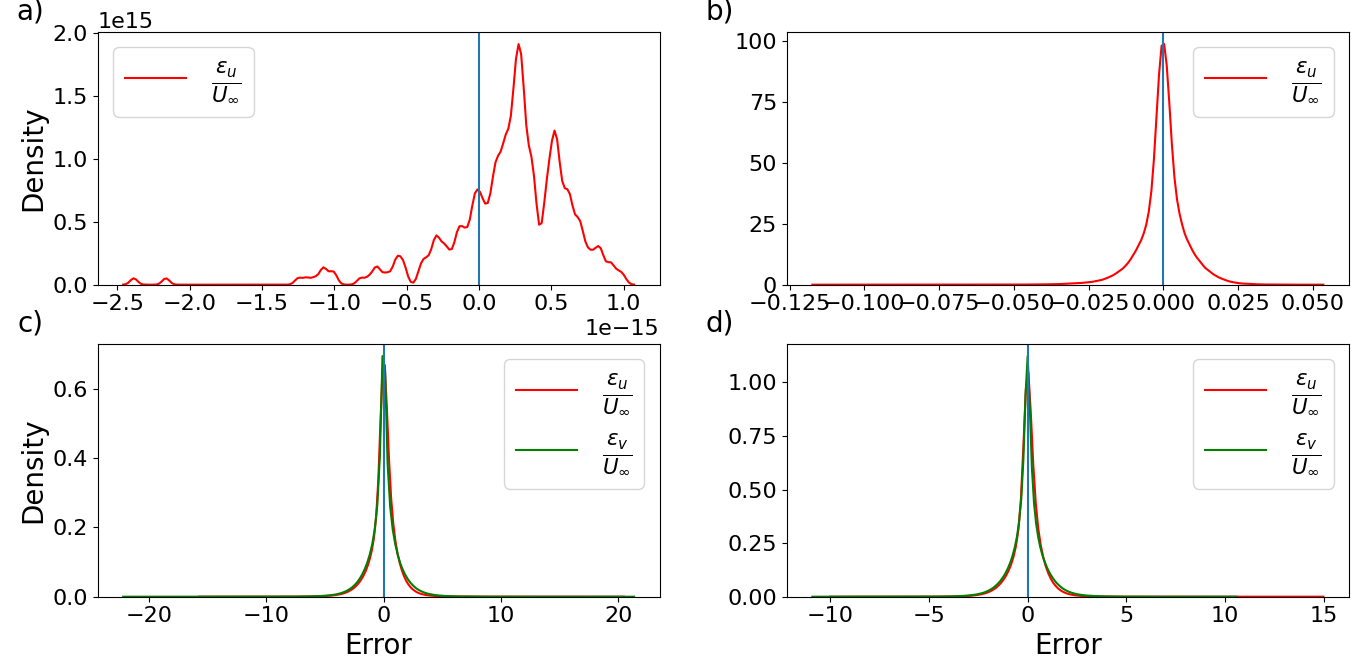}
    \caption{lcSVD reconstruction error probability density curves for: a) the streamwise velocity of the turbulent boundary layer, b) the streamwise velocity of the turbulent jet LES, c) the streamwise and normal velocity of the turbulent Re $= 4000$ cylinder, and d) the the streamwise and normal velocity of the turbulent Re $= 2600$ cylinder.}
    \label{fig:groupedPDFLCSVD}
\end{figure}

The reconstruction error probability density curves for all velocity components of the remaining these datasets are shown in Fig. \ref{fig:groupedPDFLCSVD}. 

Similar results are observed when comparing the density curves of the turbulent boundary layer, Fig. \hyperref[fig:groupedPDFLCSVD]{\ref*{fig:groupedPDFLCSVD}(a)}, and turbulent jet LES, Fig. \hyperref[fig:groupedPDFLCSVD]{\ref*{fig:groupedPDFLCSVD}(b)}, as those observed for the laminar cylinders when comparing the curves for optimal sensor and equidistant data reconstruction errors. For the boundary layer, the error values scale is reduced approximately 4 times compared to that observed in Fig. \hyperref[fig:groupedPDFOSLCSVD]{\ref*{fig:groupedPDFOSLCSVD}(a)}, and the most frequent error value is $2.5e^{-16}$, which also presents a higher density. For the turbulent jet, the left tail of the density curve expands further out, but the density of the error value 0 is higher. These results can be explained by comparing the reconstruction $RRMSE$ error values.

For the turbulent Re $= 4000$ (Fig. \hyperref[fig:groupedPDFLCSVD]{\ref*{fig:groupedPDFLCSVD}(c)}) and Re $= 2600$ (Fig. \hyperref[fig:groupedPDFLCSVD]{\ref*{fig:groupedPDFLCSVD}(d)}) cylinders the error density curves remain identical to those observed in Figs. \hyperref[fig:groupedPDFOSLCSVD]{\ref*{fig:groupedPDFOSLCSVD}(c)}, \hyperref[fig:groupedPDFOSLCSVD]{\ref*{fig:groupedPDFOSLCSVD}(d)}. Given the large error value scale, small changes in this are negligible. 

For all cases, the reconstruction error data follows the same distribution type as those visualized in the previous results subsection.

\subsection{Computational cost comparison \label{SOL_cost}}
The average computational cost for standard SVD, lcSVD and OS-lcSVD have been calculated for $N_{s} = \bar J$ = [10, 20, 30, 40, 50, 100, 125, 250] spatial data points, with the number of retained SVD modes being 10\%, 20\%, 50\%, 100\% of the number of data points for each case.

The computational cost speed-up over standard SVD is then calculated as

\begin{equation}
    S_{u}^{lcSVD} = \frac{\overline{t_{SVD}}}{\overline{t_{lcSVD}}}, \quad S_{u}^{OS-lcSVD} = \frac{\overline{t_{SVD}}}{\overline{t_{OS-lcSVD}}},
    \label{eq:speedup}
\end{equation}

where $\overline{t_{SVD}}$ is the average computational cost of SVD, $\overline{t_{lcSVD}}$ is the average cost of lcSVD, and $\overline{t_{OS-lcSVD}}$ is the average cost of OS-lcSVD measured in seconds (s). $S_{u}$ is the speed up rate referred to the SVD average computational cost. 

To calculate the average computational cost of each operation, a total of 3200 calculations have been performed for each dataset. This can be broken down into 100 iterations for each one of the 8 cases of number of selected spatial data points, and each of the 4 cases for retained SVD modes. Only 320 calculations have been performed on the turbulent boundary layer dataset, since only 10 iterations were performed instead.

Table \ref{tab:compcost} displays the average computational cost for each dataset and the respective speed-up rates when calculating for 30 data points and setting the retained SVD modes equal to 50\% of the data points.

\begin{table}[H]
\setlength{\extrarowheight}{.5ex}
    \centering
    \begin{tabular}{|l l l l l l|}
    \hline
    \rowcolor{Gray}
    \hline
    \textbf{Dataset} & \textbf{$\overline{t_{SVD}}$} (s)& \textbf{$\overline{t_{lcSVD}}$} (s)& \textbf{$\overline{t_{OS-lcSVD}}$} (s)& \textbf{$S_{u}^{lcSVD}$} & \textbf{$S_{u}^{OS-lcSVD}$}
    \\ \hline \hline
    2D Cyl100 & 3.341 & 0.47 & 2.062 & 7.109 & 1.620

    \\
    3D Cyl280 & 25.637&	4.559 &	11.761 &	5.623	&2.179

    \\
    BLayer & 1303.858 & 2.068 & 10.554 & 630.492 & 123.541

    \\
    LES jet & 62.62	&0.349	&0.984 &179.427	&63.638

    \\

    2D Cyl4000 & 101.243	&6.079	&25.772 &16.655	&3.928

    \\
    2D Cyl2600 & 204.21	&8.693	&34.357 &23.491	&5.943
    
    \\
    \hline
\end{tabular}

\caption{SVD, lc-SVD and OS-lcSVD average computational cost and speed-up rates of each algorithm and test case for $N_{s} = 30$ sensors and number of retained SVD modes equal to $50\%$ of $N_{s}$. Note that the turbulent boundary layer data has been obtained from 10 iterations. $\overline{t_{SVD}}$, $\overline{t_{lcSVD}}$ and $\overline{t_{OS-lcSVD}}$ are the average computational cost of standard SVD, lcSVD and OS-lcSVD, respectively, while $S_{u}^{lcSVD}$ and $S_{u}^{OS-lcSVD}$ are the computational cost speed-up values of lcSVD and OS-lcSVD compared to SVD, which is calculated using eq. \eqref{eq:speedup}. \label{tab:compcost}}
\end{table}

In the case of the turbulent boundary layer dataset, the average speed-up time using lcSVD on equidistant data points is around 630 times faster than using standard SVD, while the speed-up rate when using OS-lcSVD is 123 times faster. This is specially beneficial when reconstructing complex datasets, such as the laminar flow three-dimensional cylinder dataset, as well as both turbulent flow two-dimensional cylinder datasets. On the other hand, a simple dataset like the laminar flow two-dimensional cylinder presents a lower speed-up rate when applying OS-lcSVD, this being due to the already low computational cost when applying standard SVD. 
The lcSVD speed-up rate will always be higher than OS-lcSVD since this last algorithm includes the computational cost associated to the calculation of the optimal sensor positions. 

Memory consumption has also been calculated using the \textit{memory\_profiler} Python library \cite{memprof}, and the results are presented in Tab. \ref{tab:memory}

\begin{table}[H]
    \centering
    \begin{tabular}{|l l l l |}
    \hline
    \rowcolor{Gray}
    \hline
    \textbf{Dataset} & \textbf{SVD (MiB)} & \textbf{lcSVD (MiB)} & \textbf{OS-lcSVD (MiB)}
    \\ \hline \hline
    2D Cyl100 & 9924.9 & 9813.6 & 9536.2
    \\
    3D Cyl280 & 14677.0 & 13322.6 & 11340.5
    \\
    BLayer & 13980.1 & 11454.3 & 11501.1
    \\
    LES jet & 11919.3 & 11713.6 & 9384.3
    \\
    2D Cyl4000 & 17226.8 & 15368.3 & 11632.9
    \\
    2D Cyl2600 & 19438.3 & 16877.6 & 12245.7
    \\
    \hline
\end{tabular}
\caption{Memory consumption for SVD, lcSVD and OS-lcSVD, in MiB, for each dataset. \label{tab:memory}}
\end{table}

In most cases, OS-lcSVD seems to consume less memory but its execution time can be up to five times the lcSVD algorithm's execution time, as seen in Tab. \ref{tab:compcost}. Therefore, the memory consumption is more prolonged in time. 

Since a reduced amount of data is being used, the memory consumption and computational cost of lcSVD and OS-lcSVD are drastically lower than standard SVD. 

In short, lcSVD is proven to be less taxing cost- and memory-wise. OS-lcSVD is slightly more expensive than lcSVD with the time and memory increments being due to this algorithm seeking the optimal sensor positions, which is expected.

\section{Conclusions\label{sec:conclusions}}
This article has presented a new methodology named low-cost SVD, refering to its low computational cost when compared to standard singular value decomposition. 

This novel method allows for fast and precise reconstruction of fluid mechanics experimental and numerical simulation datasets using only a small amount of random or equidistant data point, or by using the data collected by a set of sensors which are optimally placed using the OS-lcSVD method. This iterative method allows users to locate the optimal sensor positions given a number of sensors. These sensor locations guarantee small reconstruction error when applying lcSVD. Users are also able to estimate the optimal number of sensors by setting an error tolerance or by applying the elbow method. This last method consists in finding the inflection point (elbow) in the reconstruction error uncertainty evolution graph for a range of available sensors.

This new methodology has been applied on a variety of datasets, and the results presented demonstrate its reduce computational cost, as well as prove its high precision when reconstructing data, revealing its strong potential to be used in big data and  fluid mechanic problems solving complex and industrial applications. The results show that, when reconstructing turbulent datasets using OS-lcSVD, with the minimum number of sensors possible, optimally positioned in the experimental domain, reconstruction errors lower than 0.5\% can be obtained with a computational cost up to 123 times lower (faster) than using standard SVD, while memory consumption can drop by almost 40\%.

The algorithms presented in this article will be incorporated into the next version release of ModelFLOWs-app, which can be downloaded from the official ModelFLOWs-app software website \cite{ModelFLOWsappWeb}.

\section[]{Acknowledgements}
The authors acknowledge fruitful discussions with Prof. Marta Cordero (Universidad Politécnica de Madrid) about uncertainty quantification and statistics. The authors acknowledge the grants PID2020-114173RB-I00, TED2021-129774B-C21 and  PLEC2022-009235 funded by MCIN/AEI/ 10.13039/501100011033 and by the European Union “NextGenerationEU”/PRTR, and S.L.C. acknolwedges the support of Comunidad de Madrid through the call Research Grants for Young Investigators from Universidad Politécnica de Madrid. 



\bibliographystyle{elsarticle-num-names} 

\end{document}